\documentclass[aps,pra,nofootinbib,onecolumn,a4paper,superscriptaddress]{revtex4-2}
\usepackage{bm,latexsym,amsmath,amssymb,amsfonts,mathrsfs,amsfonts}
\usepackage[dvipdfmx]{graphicx}
\usepackage[hang,small,bf]{caption}
\usepackage[subrefformat=parens]{subcaption}
\usepackage{ascmac}
\usepackage{physics}
\usepackage{amsmath}
\usepackage{color,float}
\usepackage{braket}
\usepackage{bbm}
\usepackage{comment}
\usepackage{here}
\usepackage{indentfirst}
\newcommand{\simeqgt}{\lower.5ex\hbox{$\; \buildrel > \over \simeq \;$}}
\newcommand{\simeqlt}{\lower.5ex\hbox{$\; \buildrel < \over \simeq \;$}}

\begin{document}
\title{Measurement-Based Estimation of Causal Conditional Variances and Its Application to Macroscopic quantum phenomenon}
\author{Kosei Hatakeyama}
\email{koseihtkym998@gmail.com}
\affiliation{Department of Physics,  Kyushu University, 744 Motooka, Nishi-Ku, Fukuoka 819-0395, Japan}
\author{Ryotaro Fukuzumi}
\email{fukuzumi.ryotaro.709@s.kyushu-u.ac.jp}
\affiliation{Department of Physics,  Kyushu University, 744 Motooka, Nishi-Ku, Fukuoka 819-0395, Japan}
\author{Akira Matsumura}
\email{matsumura.akira@phys.kyushu-u.ac.jp}
\affiliation{Department of Physics,  Kyushu University, 744 Motooka, Nishi-Ku, Fukuoka 819-0395, Japan}
\affiliation{\small\it Quantum and Spacetime Research Institute, Kyushu University, 744 Motooka, Nishi-Ku, Fukuoka 819-0395 Japan}
\author{Daisuke Miki}
\email{dmiki@caltech.edu}
\affiliation{\small\it The Division of Physics, Mathematics and Astronomy, California Institute of Technology, Pasadena, CA 91125, USA}
\author{Kazuhiro Yamamoto}
\affiliation{Department of Physics,  Kyushu University, 744 Motooka, Nishi-Ku, Fukuoka 819-0395, Japan}
\affiliation{\small\it Quantum and Spacetime Research Institute, Kyushu University, 744 Motooka, Nishi-Ku, Fukuoka 819-0395 Japan}

\begin{abstract}
We analytically investigate a quantum estimation method for a mechanical oscillator in a detuned cavity system based solely on homodyne measurement records, building on the framework developed by C.~Meng et al. (Science Advances 8, 7585 (2022)). 
Estimation based only on measurement records is important because it enables state verification without assuming knowledge of the true system state. 
We construct a relative estimate operator from causal and anti-causal quantum Wiener filters and calculate its variance. 
The deviation from the causal conditional variance is defined as a reconstruction bias, whose magnitude is evaluated analytically.
We show that, within experimentally relevant parameter regimes for typical quantum-state preparation, the reconstruction bias is sufficiently small to be neglected. As applications to state verification, we apply the method to proposals for macroscopic quantum entanglement mediated by electromagnetic interactions and for conditional momentum-squeezed states generated by homodyne detection, and clarify the conditions under which the bias remains negligible and when the reconstruction bias becomes significant.
\end{abstract}
\maketitle

\section{Introduction}
Optomechanical systems provide a versatile platform for exploring quantum phenomena by controlling mechanical oscillators with light, enabling ground-state cooling, the preparation of nonclassical states, and the generation of mechanical-mechanical or optomechanical entanglement \cite{Chen2013,Aspelmeyer2014}. Two widely used approaches to access a mechanical oscillator's quantum state are pulsed optical tomography \cite{Vanner2011,Hofer2011} and conditional state preparation based on continuous measurement records \cite{Wiener1949,Muller2008,Muller2009,Bouten2004,Yamamoto2006,Wieczorek2015}. In the former, by appropriately choosing the cavity detuning, one can implement an effective state-swap interaction that maps the mechanical state onto an optical mode, which can be measured. Experimentally, this technique has been used, for instance, to detect entanglement between two cooled mechanical oscillators \cite{Korppi2018,Kotler2021,Lepinay2021}. 
In the latter approach, the oscillator is continuously monitored via the measurement of the output optical light; the mechanical state is conditioned on the measurement outcomes and is described by a conditional quantum state obtained through a causal estimation procedure. This method does not require additional probe pulses beyond the measurement already used for control since the measurement record is processed to infer the conditional state. Near the quantum regime, conditional mechanical squeezing has been demonstrated experimentally \cite{Matsumoto}, and such conditional states are expected to enable entanglement between two macroscopic oscillators in future experiments \cite{Miki2023}.

We consider the verification of conditional mechanical states from continuous measurement records.
For linear Gaussian dynamics and measurements, the conditional state is fully characterized by its conditional covariance matrix.
However, experimentally verifying this covariance matrix is not straightforward, since it is usually defined in terms of the unconditional covariance, which is not directly accessible from the measurement record alone.
Although Wiener-filter-based estimation assumes a linear Gaussian model, the unconditional covariance must still be supplied separately, for example through calibration or by fitting the measured output spectra.
Motivated by this limitation, previous studies and the present work seek to derive the conditional covariance directly from the measurement record, without explicitly relying on the unconditional covariance.
To address this issue, Refs.~\cite{Rossi,Matsumoto,Meng2022} proposed a verification strategy that combines causal estimation with time-reversed estimation (retrodiction). 
The key idea is to construct two independent estimators for the same mechanical variables using non-overlapping portions of the measurement record: a causal estimator using measurement data prior to a given time and an anti-causal estimator using data obtained after that time.
By comparing the statistics of these two estimators, one can experimentally assess the consistency of the conditional state without explicitly relying on theoretical knowledge of the unconditional covariance.

In Ref.~\cite{Meng2022}, a verification scheme for mechanically conditioned Gaussian states was formulated in the Fourier domain using Wiener filtering for a double-disk optomechanical model. The key idea is to reconstruct the conditional variances using only the experimentally available measurement record by combining a causal (prediction) Wiener filter with an anti-causal (retrodiction) one, thereby avoiding any dependence on unobservable “true” trajectories. In that work, the analysis was restricted to optical phase-quadrature measurements and a tuned cavity.

Motivated by this record-only reconstruction approach, we generalize the same measurement-record–based estimation/verification scheme to detuned optomechanical systems with an arbitrary homodyne angle. In general, the homodyne angle and cavity detuning are closely tied to the mechanical state being prepared; for example, Ref.~\cite{Fukuzumi2025} showed that appropriate parameter choices can realize a momentum-squeezed conditional state, which can enhance the entanglement generated by position-correlated interactions.

In the present paper, we develop an analytical framework for measurement-based estimation in a detuned cavity–mechanical oscillator system under an arbitrary homodyne angle. We formulate the anti-causal filter following Meng's method and combine it with the causal Wiener filter, which enables the reconstruction of conditional variances solely from the estimators.
We show that the resulting variances can be written as the sum of the causal conditional variances and a reconstruction bias, and derive an analytical expression for this bias.
Numerical results confirm that this bias is sufficiently small to be neglected over a broad range of a mechanical dissipation rate and input optical power for almost all homodyne angles.
In particular, we derive simple approximate expressions for the reconstruction bias for measurements of the amplitude and phase quadratures, showing that a high mechanical quality factor and strong cooperativity effectively suppress it.
In applications, we analyze macroscopic quantum entanglement mediated by electromagnetic interactions \cite{Miki2023} and conditional momentum-squeezed states generated by homodyne detection \cite{Fukuzumi2025}. For entanglement verification, we show that the reconstruction bias is negligible within previously studied parameter regimes. In contrast, we derive an approximate expression for the bias in measurement settings where momentum squeezing is pronounced and find that, at zero detuning, it increases significantly with laser power. These results indicate that in the high-power regime, both the choice of detuning and the reconstruction bias must be carefully taken into account in a quantitative evaluation of momentum-squeezed states.

This paper will be organized as follows.
Section II reviews the basic formulation of the cavity optomechanical system.
Section III introduces the causal and anti-causal Wiener filters. We analytically examine the measurement-based estimation scheme proposed by Rossi et al. \cite{Rossi} and Meng et al. \cite{Meng2022} and formulate the reconstruction bias.
Section IV applies the measurement-based estimation to re-evaluate macroscopic quantum entanglement mediated by electromagnetic interactions \cite{Miki2023} and momentum-squeezed states generated by homodyne detection \cite{Fukuzumi2025}.
Section V presents the conclusions and outlook.
Appendix A summarizes the definitions of the causal and anti-causal Wiener filters, and Appendix B lists the integral formulas used in the calculation of the reconstruction bias.

\section{Basic Model}
In this section, we briefly review the dynamics of the optomechanical system in this work. 
The original system consists of a 
mechanically oscillating mirror (mass $m$ and resonance frequency $\Omega$) and an optical cavity (length $\ell$ and cavity resonance frequency $\omega_c$), which has been extensively discussed in previous studies (e.g., \cite{Matsumoto, Miki2024a,Miki2024b,Miao2020, Datta2021, Fukuzumi2025, Hatakeyama2025}).  The notation and normalization conventions adopted in this work follow Ref.~\cite{Fukuzumi2025}.
The total Hamiltonian is given by:
\begin{align}
    \hat{H}&=\frac{\hat{P}^2}{2m}+\frac{m\Omega^2\hat{Q}^2}{2}+\hbar\omega_c\hat{a}^\dagger\hat{a}+\frac{\hbar\omega_c}{\ell}\hat{Q}\hat{a}^\dagger\hat{a}+\frac{i\hbar E}{\sqrt{2}}\Big(e^{-i\omega_L t}\hat{a}^\dagger-e^{i\omega_L t}\hat{a}\Big),\label{Hamiltonian}
\end{align}
where $\hat{Q}$ and $\hat{P}$ are the position and momentum operators of the mirror, satisfying $[\hat{Q},\hat{P}]=i\hbar$.
$\hat{a}$ and $\hat{a}^{\dagger}$ are the annihilation and creation operators of optical cavity modes, with $[\hat{a},\hat{a}^{\dagger}]=1$. 
$E=\sqrt{P_{\rm in}\kappa/\hbar\omega_L}$ is the laser amplitude, where $P_{\rm in}$ is the input laser power, and $\kappa$ is the cavity decay rate.
Hereafter, in the numerical calculations, we set $\ell = 10^{-1}$[m] and $\omega_c \simeq \omega_L = 2\pi \times 2.818 \times 10^{14}$[Hz].
The terms on the right-hand side  of Eq.~\eqref{Hamiltonian} represent the Hamiltonian for the mirror and the cavity field, as well as the coupling between them via radiation pressure \cite{Aspelmeyer2014}. 

We consider the perturbations around a steady state by writing $\hat{Q}\rightarrow\bar{Q}+\hat{Q}$, $\hat{P}\rightarrow\bar{P}+\hat{P}$, $\hat{a}\rightarrow\bar{a}+\hat{a}$, $\hat{a}^{\dagger}\rightarrow\bar{a}^{*}+\hat{a}^{\dagger}$, where $(\bar{Q}, \bar{P} ,\bar{a})$ are the classical mean values, while  $(\hat{Q}, \hat{P} ,\hat{a})$ are the perturbative quantities.
By introducing the mechanical frequency modified by optical spring effect $\omega_m=\sqrt{\Omega^2+16\Delta g^2\Omega/(\kappa^2+4\Delta^2)}$ and dimensionless operators $\hat{q}=\sqrt{2m\omega_m/\hbar}\hat{Q}, \hat{p}=\sqrt{2/(m\hbar \omega_m)}\hat{P}$, the equations of motion for the perturbative quantities are given as follows \cite{Matsumoto, Fukuzumi2025}:
\begin{align}
    \dot{\hat{q}}&=\omega_m\hat{p}, \label{EoMq}\\
    \dot{\hat{p}}&=-\omega_m\hat{q}-\Gamma\hat{p}+\sqrt{2\Gamma}p_\text{in}
    -\sqrt{\omega_m\xi}(x_\text{in}\cos\alpha-y_\text{in}\sin\alpha),\label{EoM}\\
   \dot{\hat{x}} &= -\Delta \hat{y} - \frac{\kappa}{2}\hat{x}
    + \sqrt{\kappa}x_\mathrm{in}, \\
    \dot{\hat{y}} &= \Delta \hat{x} - 2 g_m \hat{q}
    - \frac{\kappa}{2}\hat{y}
    + \sqrt{\kappa}y_\mathrm{in}, \\
    \xi&=\frac{16g_m^2}{\kappa\omega_m}\cos^2\alpha,
\end{align}
where $g_m=\sqrt{\omega_cP_\text{in}\kappa/(m\omega_m\ell^2(\kappa^2+4\Delta^2))}$ is the optomechanical coupling constant, $\Gamma$ is the mechanical damping rate, and $\kappa$ is the optical decay rate in the cavity field.
$\alpha=\arctan{(2\Delta/\kappa)}$ represents the phase acquired by the input laser upon entering the cavity.
The correlation function of the input thermal noise satisfies $\langle\{p_\mathrm{in}(t), p_\mathrm{in}(t')\}\rangle = 2(2n_\mathrm{th} + 1)\delta(t-t')$, where the thermal phonon number is given by $n_\text{th} = k_B T / (\hbar \omega_m)$. Throughout this work, the temperature is uniformly set to room temperature, $T = 300\mathrm{K}$.
$\xi$ corresponds to the magnitude of the  radiation-pressure noise normalized by $\omega_m$. 
Similarly, the two point correlation of the input optical noise is expressed as $\langle\{x_\mathrm{in}(t), x_\mathrm{in}(t')\}\rangle = \langle\{y_\mathrm{in}(t), y_\mathrm{in}(t')\}\rangle = 2(2N_\mathrm{th}+1)\delta(t-t')
$ and $N_\text{th}=k_BT/\hbar\omega_c$ is the thermal photon number. 
The commutation relation holds $[\hat{q},\hat{p}]=2i$.
This renormalized thermal noise can be reduced by increasing $\omega_m$, and is suppressed in the regime where the optical spring effect is strong.

Next, we discuss the formulation for the output light from the cavity.
Under the adiabatic approximation $(\kappa\gg\Omega)$, which allows for the continuous measurement of the oscillator position because the cavity photon dissipation is sufficiently larger than the frequency of the oscillator, we assume $\dot{\hat{x}}=\dot{\hat{y}}=0$. 
We assume perfect homodyne measurement with angle $\theta$. Following the input-output relation \cite{Gardiner1985,Aspelmeyer2014,Fukuzumi2025}, we obtain 
\begin{align}
    x_\text{out}=x_\text{in}-\sqrt{\kappa}\hat{x}, \quad y_\text{out}=y_\text{in}-\sqrt{\kappa}\hat{y}.
\end{align}
We consider performing homodyne detection on the above output quadrature by mixing it at a homodyne angle $\theta$. 
The corresponding measurement operator is given by $\hat{I}_\theta =x_\text{out}\cos{\theta}+y_\text{out}\sin\theta =c_\theta\hat{q}+\hat{v}_\theta.$
The gain of the homodyne measurement is represented as $c_\theta=\sqrt{\omega_m\xi}\sin(\theta-\alpha)$, and $\hat{v}_\theta=-x_\mathrm{in}\cos(2\alpha-\theta)+y_\mathrm{in}\sin(2\alpha-\theta)$ represents the shot noise operators. 
The magnitude of the shot noise is independent of the homodyne angle and is given by $\langle\{\hat{v}_\theta(t), \hat{v}_\theta^\dagger(t')\}\rangle = 2(2N_\mathrm{th}+1)\delta(t-t')$.

Defining the Fourier transform $\displaystyle{f(\omega)=\int^\infty_{-\infty}dtf(t)e^{i\omega t}}$, we can obtain the solutions of the equations of motion \eqref{EoMq} and \eqref{EoM} in the frequency domain as follow;
\begin{align}
   &\hat{q}(\omega)=\frac{\omega_m\hat{w}(\omega)}{F(\omega)}, \quad \hat{p}(\omega)=-\frac{i\omega}{\omega_m}\hat{q}(\omega),\quad \hat{I}_\theta(\omega)=c_\theta\hat{q}(\omega)+\hat{v}_\theta\label{SolEOM}\\
&F(\omega)=\omega_m^2-i\Gamma\omega-\omega^2,\quad \hat{w}(\omega)=\sqrt{2\Gamma}p_\text{in}(\omega)-\sqrt{\omega_m\xi}(x_\text{in}(\omega)\cos\alpha-y_\text{in}(\omega)\sin\alpha).
\end{align}
$F(\omega)^{-1}$ represents the mirror's susceptibility, and $\hat{w}(\omega)$ is the sum of the operators of thermal noise and radiation noise of the mirror.
\vskip\baselineskip
Next, we introduce the spectral density and cross spectral density. The definition of the spectral density is given as $2\pi\delta(\omega-\omega') S_{AB}=\langle\{\hat{A}(\omega),\hat{B}^\dagger(\omega')\}\rangle/2$. 
Using the above solution, the spectral density of the mechanical oscillator, that of the output field, and the cross spectrum between them are obtained as follows.
\begin{align}
    S_{qq}(\omega)&=\frac{\omega_m^2\bar{n}}{|F(\omega)|^2}, \quad S_{pp}(\omega)=\frac{\omega^2}{\omega_m^2}S_{qq}(\omega),\quad S_{qp}(\omega)=\frac{i\omega}{\omega_m}S_{qq}(\omega),\label{unconditional spectrum}\\
    S_{II}(\omega)&=c_\theta^2S_{qq}(\omega)+c_\theta\text{Re}[S_{qI}(\omega)]+ M_\theta,\\
    S_{qI}(\omega)&=c_\theta S_{qq}(\omega)+\frac{\omega_mL_\theta}{F(\omega)}, \quad S_{pI}(\omega)=-\frac{i\omega}{\omega_m}S_{qI}(\omega),
\end{align}
where we defined
\begin{align}
    \bar{n}&=S_{ww}(\omega)=2\Gamma(2n_\text{th}+1)+\omega_m\xi
    ,\quad M_\theta=S_{vv}(\omega)=2N_\text{th}+1,\qquad
    L_\theta=S_{wv}(\omega)=
    \sqrt{\omega_m\xi}\cos{(\theta-\alpha)}.
\end{align}
where we assume that the noise correlation functions in frequency domain are given by $\langle\{p_\text{in}(\omega),p_\text{in}(\omega')\}\rangle=2(2n_\text{th}+1)\times2\pi\delta(\omega+\omega')$ and $\langle\{x_\text{in}(\omega),x_\text{in}(\omega')\}\rangle=\langle\{y_\text{in}(\omega),y_\text{in}(\omega')\}\rangle=2(2N_\text{th}+1)\times2\pi\delta(\omega+\omega')$ \cite{GardinerZoller2010QuantumNoise}. 
$\bar{n}$ represents the total noise acting on the mirror, where the first term corresponds to thermal noise and the second term represents the radiation-pressure backaction induced by the laser field. The quantity $M_\theta$ denotes the shot noise of the measurement. In the regime $k_B T \gg \hbar \omega_c$, the thermal occupation number can be approximated as $N_{\mathrm{th}} \simeq 0$, and therefore we treat $M_\theta \simeq 1$ throughout the following analysis.
The quantity $L_\theta$ is the cross spectrum between the noise acting on the mirror and the shot noise, and it characterizes the correlation between radiation-pressure backaction and measurement shot noise. In the case of phase-quadrature detection ($\theta=\pi/2$) with zero detuning ($\Delta = 0$), this correlation vanishes, and we have $L_\theta = 0$.

The equations in \eqref{unconditional spectrum} represent the unconditional spectrum of the system, which does not include the effect of measurement. 
By integrating this spectrum over the entire frequency domain, one obtains the unconditional variances,
\begin{align}
    V_{q}&=\int^\infty_{-\infty}\frac{d\omega}{2\pi}S_{qq}(\omega)=\frac{\bar{n}}{2\Gamma},\qquad V_{p}=\int^\infty_{-\infty}\frac{d\omega}{2\pi}S_{pp}(\omega)=\frac{\bar{n}}{2\Gamma},\qquad
    V_{qp}=\int^\infty_{-\infty}\frac{d\omega}{2\pi}S_{qp}(\omega)=0.
\end{align}
While these quantities can be evaluated theoretically, they cannot be determined directly from experimentally accessible measurement records alone.

\section{measurement-based estimation of the covariance in Fourier domain}
In this section, we construct a framework for quantum optimal estimation based on causal and anti-causal Wiener filtering. Conventionally, quantum causal estimation in the frequency domain focuses on inferring the current system state solely from past measurement records. Within this framework, the estimation performance is characterized by the variance of the deviation between the true system operators, $\hat{q}(\omega)$ and $\hat{p}(\omega)$, and their optimal causal estimates, which is referred to as the (causal) conditional variance. This quantity represents the uncertainty that remains in the system after optimal estimation conditioned on the measurement quantities.

However, this formulation cannot be directly applied in experiments, since the true system operators $(\hat{q}(\omega), \hat{p}(\omega))$ are fundamentally inaccessible without being affected by measurement backaction and additional noise. To overcome this limitation, Meng et al. adopted an approach that combines causal and anti-causal Wiener filters to reconstruct the causal conditional variance without relying on the true system values \cite{Meng2022}.
This method is based on the decomposition of the relevant spectra into causal and anti-causal components and enables the evaluation of conditional variances using only experimentally accessible information. Following the approach employed by Meng et al., we formulate the method for a Fabry–Perot cavity system consisting of a movable mirror coupled to an optical cavity, and investigate its applicability to evaluating the purity and quantum entanglement between macroscopic objects.

Related studies include~\cite{Matsumoto,Rossi}. In particular, \cite{Matsumoto} considered an optomechanical system similar to ours and evaluated the causal conditional variance without directly using the unconditional variance by combining predictive and retrodictive equations of motion with Kalman filtering in the time domain. Their method is theoretically sound and has been experimentally validated. Although their approach does not follow the same analytical procedure as that of Meng et al., it constitutes an important result. In the present work, we adopt the method of Meng et al., which allows for a relatively simple construction of the anti-causal Wiener filter, and base the following analysis on this framework.

We first briefly review causal estimation based on the quantum Wiener filter under homodyne measurement. In this framework, the system observables are estimated from the continuous measurement record under the constraint of causality, which in the frequency domain is implemented by restricting the filter functions to causal transfer functions. The quantum Wiener filter provides a systematic way to construct optimal estimators by minimizing the mean-square error between the system operators and their estimation operators, while properly accounting for quantum measurement noise and backaction.

In causal estimation, the conditional variances characterize the residual uncertainty of the system under continuous measurement. They are defined as the unconditional variances of the system operators minus the variances of their causal estimation values and represent the quantum uncertainty that cannot be reduced by any causal data processing.
The causal estimation values are defined as
$\overrightarrow{q}(\omega)=\overrightarrow{H}_{q}(\omega)\hat{I}_\theta(\omega)$ and
$\overrightarrow{p}(\omega)=\overrightarrow{H}_{p}(\omega)\hat{I}_\theta(\omega)$,
where $\overrightarrow{H}_{q}(\omega)$ and $\overrightarrow{H}_{p}(\omega)$ are causal quantum Wiener filters constructed to optimally estimate the mechanical position and momentum from the measurement record using only past information (see Appendix~A).
\begin{align}
    \overrightarrow{H}_q(\omega)&:=\frac{1}{S_{II}^\text{C}(\omega)}\Bigg[\frac{S_{qI}(\omega)}{S_{II}^\text{AC}(\omega)}\Bigg]_+=\frac{\omega_\theta^2-\omega_m^2-i\omega(\gamma_\theta-\Gamma)}{c_\theta F'(\omega)}, \label{causal-filter-q}\\
    \overrightarrow{H}_p(\omega)&:=\frac{1}{S_{II}^\text{C}(\omega)}\Bigg[\frac{S_{pI}(\omega)}{S_{II}^\text{AC}(\omega)}\Bigg]_+=\frac{-(\gamma_\theta-\Gamma)\omega_m^2-i\omega(\omega_\theta^2-\omega_m^2-(\gamma_\theta-\Gamma)\Gamma)}{c_\theta\omega_m F'(\omega)}, \label{causal-filter-p}
\end{align}
where $S_{II}^\text{C}(\omega)$ and $S_{II}^\text{AC}(\omega)$ denote the causal and anti-causal factors of $S_{II}(\omega)$, respectively.  These satisfy the relations $S_{II}^\text{C}(\omega)S_{II}^\text{AC}(\omega)=S_{II}(\omega)$ and $S_{II}^\text{C}(\omega)=(S_{II}^\text{AC}(\omega))^*$. The symbols
$[\cdots]_+$ and $[\cdots]_-$ denote the operations that extract the causal and anti-causal components, respectively \cite{Shichijo,kisil2021wienerhopf,Wiener1949}.
These filters are identical to those introduced in Ref.~\cite{Fukuzumi2025}, and in the cases of $X$ and $Y$ quadrature measurements, they coincide with the results reported in Ref.~\cite{Matsumoto}.
In this convention, a function is called causal if it satisfies $f(t)=0$ for $t<0$. In this case, the corresponding frequency-domain function is analytic in the upper half of the complex frequency plane and has poles only in the lower half-plane.
Conversely, an anti-causal function satisfies $f(t)=0$ for $t>0$, and its poles lie in the upper half-plane.
Here, $\omega_\theta$ and $\gamma_\theta$ are the effective mechanical frequency and dissipation rate under estimation, respectively, and $F'(\omega)^{-1}=(\omega_\theta^2-i\gamma_\theta\omega-\omega^2)^{-1}$ represents the effective susceptibility of the mirror in the estimation~\cite{Matsumoto,Fukuzumi2025}.
The explicit expressions of $\omega_\theta$ and $\gamma_\theta$ are as follows:
\begin{align}
    \omega_\theta&=\sqrt[4]{\omega_m^4+2\Lambda_\theta\omega_m^3+\bar{n}\lambda_\theta\omega_m^2},\quad\gamma_\theta=\sqrt{\Gamma^2-2\omega_m(\omega_m+\Lambda_\theta)+2\omega_\theta^2},\\
    \lambda_\theta&=\frac{c_\theta^2}{M_\theta}=\omega_m\xi\sin^2(\theta-\alpha)
    , \quad \Lambda_\theta=\frac{c_\theta L_\theta}{M_\theta}=\omega_m\xi\sin(\theta-\alpha)\cos{(\theta-\alpha)}.
\end{align}
Here, $\lambda_\theta$ denotes the homodyne measurement strength, which characterizes the rate of information acquisition in the continuous measurement process. Equivalently, it sets the inverse timescale required to resolve the mechanical zero-point fluctuations.
% $M_\theta=\langle(\hat{v}_\theta)^2\rangle=2N_\text{th}+1$ is the magnitude of the shot noise.
% $L_\theta=\langle\hat{w}\hat{v}_\theta\rangle$ denotes the cross-correlation between the radiation-pressure noise acting on the oscillator and the shot noise.

We obtain the causal covariance matrix $V=\begin{pmatrix}
V_{q}^\text{c} & V_{qp}^\text{c} \\
V_{qp}^\text{c} & V_{p}^\text{c}
\end{pmatrix} $ by integrating the conditional spectrum over the entire frequency domain. 
More explicitly, integrating the causal conditional spectral densities yields
\begin{align}
    V_{q}^\text{c}&=\int^\infty_{-\infty}\frac{d\omega}{2\pi}\Big(S_{qq}(\omega)-S_{\overrightarrow{q}\overrightarrow{q}}(\omega)\Big)=V_{q}-V_{\overrightarrow{q}}=\frac{\gamma_\theta-\Gamma}{\lambda_\theta},\\
    V_{p}^\text{c}&=\int^\infty_{-\infty}\frac{d\omega}{2\pi}\Big(S_{pp}(\omega)-S_{\overrightarrow{p}\overrightarrow{p}}(\omega)\Big)=V_{p}-V_{\overrightarrow{p}}=\frac{\gamma_\theta-\Gamma}{\lambda_\theta}\Bigg[\Big(\frac{\omega_\theta}{\omega_m}\Big)^2-\frac{\gamma(\gamma_\theta-\Gamma)}{2\omega_m^2}\Bigg],\label{causal-V}\\
    V_{qp}^\text{c}&=\int^\infty_{-\infty}\frac{d\omega}{2\pi}\text{Re}\Big(S_{qp}(\omega)-S_{\overrightarrow{q}\overrightarrow{p}}(\omega)\Big)=V_{qp}-V_{\overrightarrow{q}\overrightarrow{p}}=\frac{(\gamma_\theta-\Gamma)^2}{2\lambda_\theta\omega_m}\label{causal-Vqp}.
\end{align}
As discussed above, the conditional variances are expressed in terms of the unconditional variances. However, these quantities are not directly accessible experimentally. In quantum measurements, there are no underlying “true values” of system observables that are merely hidden by noise. Rather, the intrinsic dynamics of the system cannot be accessed independently of the measurement process, since shot noise inevitably accompanies any measurement. Consequently, a direct experimental verification of the causal estimation scheme would require the unconditional variances to be provided as theoretically determined inputs.

However, merely providing such theoretical values does not in itself constitute an independent experimental verification, since the any resulting comparison inevitably relies on the assumed consistency between theory and measurement. For a more direct assessment of the causal estimation scheme, it is therefore desirable to infer the causal conditional variances without explicit reference to the true system operators or to externally supplied unconditional variances.

This can be achieved by combining causal and anti-causal estimators constructed from the same measurement record. In this fremework, the causal estimator describes the state inferred from past measurement outcomes, whereas the anti-causal estimator is obtained by estimating the state using measurement data extending into later times. By jointly employing these two estimators, the causal conditional variances can be reconstructed without reference to the unobservable true system operators. 
The anti-causal estimation is not intended for real-time state monitoring, in contrast to Kalman or causal Wiener filters. Rather, it is implemented as a post-processing procedure once the measurement has been completed, and serves as an essential complement to the causal estimation.
\begin{align}
\overleftarrow{H}_q(\omega)&=\frac{1}{S_{II}^\text{AC}(\omega)}\Bigg[\frac{S_{qI}^*(\omega)}{S_{II}^\text{C}(\omega)}\Bigg]_-=\Big(\overrightarrow{H}_q(\omega)\Big)^*,\quad \overleftarrow{H}_p(\omega)=\frac{1}{S_{II}^\text{AC}(\omega)}\Bigg[\frac{S_{pI}^*(\omega)}{S_{II}^\text{C}(\omega)}\Bigg]_-=-\Big(\overrightarrow{H}_p(\omega)\Big)^*.\label{anti-causal_filter}
\end{align}
Details of the derivation of the causal and anti-causal Wiener filters defined in Eqs.~\eqref{causal-filter-q}, \eqref{causal-filter-p}, and \eqref{anti-causal_filter} are provided in Appendix A.

Meng et al. \cite{Meng2022} introduced relative-estimate operators constructed by combining causal and anti-causal filters and demonstrated that the variance of this estimator reproduces, to good accuracy, the conditional variances obtained from causal estimation, $(V_q^{\mathrm{c}}, V_p^{\mathrm{c}})$.
In the present work, we refer to the approach introduced by Meng et al. as measurement-based estimation.
Following this approach, we define a relative-conditional operator combining both the causal and anti-causal filters with the measurement operator as follows:
\begin{align}
    \Delta q(\omega)&
    =\frac{\overrightarrow{H}_q(\omega)-\overleftarrow{H}_q(\omega)}{\sqrt{2}}\hat{I}_\theta(\omega),\quad 
    \Delta p(\omega)
    =\frac{\overrightarrow{H}_p(\omega)-\overleftarrow{H}_p(\omega)}{\sqrt{2}}\hat{I}_\theta(\omega).\label{relative-operator}
\end{align}
A difference between the variances of these operators defined in Eq.~\eqref{relative-operator} and the conditional variances obtained from causal estimation arises when the system exhibits dissipation. The corresponding spectral densities can then be written as follows:
\begin{align}
    S_{\Delta q \Delta q}(\omega)=\frac{\Big|\overrightarrow{H}_q(\omega)-\overleftarrow{H}_q(\omega)\Big|^2}{2}S_{II}(\omega), \qquad S_{\Delta p \Delta p}(\omega)=\frac{\Big|\overrightarrow{H}_p(\omega)-\overleftarrow{H}_p(\omega)\Big|^2}{2}S_{II}(\omega).
\end{align}
The above quantities are expressed entirely in terms of the two filter functions and the output optical spectrum associated with the measurement record. They are thus defined exclusively using experimentally accessible quantities.
By analytically performing the frequency integrals, we explicitly evaluate this discrepancy and express the deviation from the causal conditional variances as follows:
\begin{align}
     &V_{\Delta q}=\int^{\infty}_{-\infty}\frac{d\omega}{2\pi}S_{\Delta q\Delta q}(\omega)
     =V_{q}^\text{c}+\alpha_\theta, \qquad   V_{\Delta p}=
     \int^{\infty}_{-\infty}\frac{d\omega}{2\pi} S_{\Delta p\Delta p}(\omega)
     =V_{p}^\text{c}+\beta_\theta,\\
     &\alpha_\theta=-\frac{2\Gamma\omega_m}{c_\theta G^2}\Bigg(c_\theta\omega_m\Big(\omega^{2}_m(\gamma_\theta+\Gamma)^2-(\omega^{2}_\theta-\omega^{2}_m)^2\Big)\bar{n}+2\Gamma(\omega^{2}_\theta-\omega^{2}_m)\Big(2\gamma_\theta\omega_m^2+\Gamma(\omega_\theta^2+\omega_m^2)\Big)L_\theta\Bigg),\label{alpha}\\
     &\beta_\theta=\frac{\Gamma}{c_\theta\omega_mG^2}\nonumber\\
     &\times\Bigg[2c_\theta\omega_m\Bigg(\omega^{6}_m+\Big(2(\gamma^{2}_\theta-\omega^{2}_m)-3\Gamma^2\Big)\omega^{4}_m+\Big(\omega^{4}_\theta-2(\gamma^{2}_\theta+\gamma_\theta\Gamma-\Gamma^2)\omega^{2}_\theta+(\gamma^{2}_\theta-\Gamma^2)^2\Big)\omega^{2}_m-\gamma^{2}_\theta
     \omega^{4}_\theta\Bigg)\bar{n}\nonumber\\
     &+\Bigg(\omega^{8}_m+2(\gamma^{2}_\theta+\gamma^{}_\theta\Gamma-4\Gamma^2-2\omega^{2}_\theta)\omega^{6}_m+\Big(6\omega^{4}_\theta-2(2\gamma^{2}_\theta+5\Gamma\gamma_\theta-5\Gamma^2)\omega^{2}_\theta+\gamma^{2}_\theta(\gamma^{2}_\theta+2\gamma_\theta\Gamma-7\Gamma^2)+4\Gamma^4\Big)\omega^{4}_m\nonumber\\
     &\qquad+2\omega^{2}_\theta\Big(-\Gamma\gamma_\theta(3\gamma_\theta+\Gamma)(\gamma^{}_\theta-\Gamma)+(\gamma^{2}_\theta+3\Gamma\gamma^{}_\theta-2\Gamma^2-2\omega^{2}_\theta)\omega^{2}_\theta\Big)\omega^{2}_m\nonumber\\
     &\qquad+\omega^{4}_\theta(\omega^{2}_\theta+\Gamma^2-\Gamma\gamma^{}_\theta)(\omega^{2}_\theta+\Gamma^2+3\Gamma\gamma_\theta)\Bigg)L_\theta\Bigg],\label{beta}\\
     &V_{\Delta q\Delta p}=\int^{\infty}_{-\infty}\frac{d\omega}{2\pi} S_{\Delta q\Delta p}(\omega)=0,\label{reCV}
\end{align}
where we defined $G=\gamma_\theta^2\omega_m^2+\Gamma^2\omega_\theta^2+(\omega_\theta^2-\omega_m^2)^2+\gamma_\theta\Gamma(\omega_\theta^2+\omega_m^2)$. 
Here, $\alpha_\theta$ and $\beta_\theta$ correspond to the reconstruction bias for position and momentum, respectively. This result demonstrates the variances obtained through this estimation method--without access to the true system values--deviate from the causal conditional variances by $\alpha_\theta$ and $\beta_\theta$. 
These reconstruction bias can be neglected when the dissipation rate $\Gamma$ vanishes.
Details of these integral calculations are provided in Appendix B.
The relative estimated covariance $V_{\Delta q\Delta p}$ vanishes because the causal covariance $V_{qp}$ and the anti-causal covariance $V_{qp}^\text{AC}$ have opposite signs and cancel each other. Although this method does not directly reproduce the causal estimation covariance, $V_{qp}$ can nevertheless be computed without access to the true system values, as noted above. We therefore proceed with the discussion using $V_{qp}$.

We investigate how the reconstruction bias can be approximated when thermal noise is sufficiently small ($\bar{n}\simeq\omega_m\xi$). 
For phase measurement ($\theta = \pi/2$) in the regime $\Delta = 0$ and $\Gamma \ll \Omega$, one may use the approximations $\omega_\theta \simeq 4g\sqrt{\Omega/\kappa}$ and $\gamma_\theta \simeq \sqrt{2}\omega_\theta$, which yield
\begin{align}
    \alpha_\frac{\pi}{2}\simeq\frac{1}{2C},\qquad \beta_\frac{\pi}{2}\simeq-\frac{4}{Q} \label{bias-Y},
\end{align}
where $C=4g^2/(\Gamma\kappa)$ is the cooperativity, and $Q=\Omega/\Gamma$ is the mechanical quality factor. Therefore, increasing the laser power reduces the reconstruction bias of the relative-estimate variance for the mirror position, whereas decreasing the mechanical dissipation reduces the reconstruction bias in the relative-estimate variance for the mirror momentum.

On the other hand, in the case of amplitude measurement ($\theta=0$), a nonzero detuning is required for the measurement, and the corresponding approximation differ from those used in the phase-measurement case. We focus on the regime of sufficiently large laser power $P_\text{in}$, 
where $\Gamma\ll\gamma_\theta,~\Omega$ and $\kappa^2\gg\Delta^2$. 
In this regime, one may approximate $\omega_\theta\simeq\omega_m$ and $\gamma_\theta\simeq\sqrt{2}\omega_m\sqrt{\xi\sin{\alpha}\cos{\alpha}}$. 
Furthermore, even for small detuning, sufficiently strong laser driving allows the use of the relation $\xi\simeq\kappa/\Delta$, which implies $\omega_m/\gamma_\theta^2\simeq\Delta/(c_\theta^2\kappa)\simeq1/(4\omega_m)$. 
We then find the approximated expression,
\begin{align}
    \alpha_0\simeq\frac{\xi}{2Q_m},\qquad \beta_0\simeq-\frac{\xi}{Q_m} ,\label{bias-X}
\end{align}
where $Q_m=\omega_m/\Gamma$ is the mechanical quality factor normalized by $\omega_m$.
In this case, the normalized radiation-pressure noise $\xi$ becomes a constant that no longer depends on the laser power. Therefore, in amplitude-quadrature measurements, increasing the mechanical quality factor is essential for reducing the reconstruction bias.
\begin{figure}[t]
\begin{center}
\includegraphics[width=86mm]{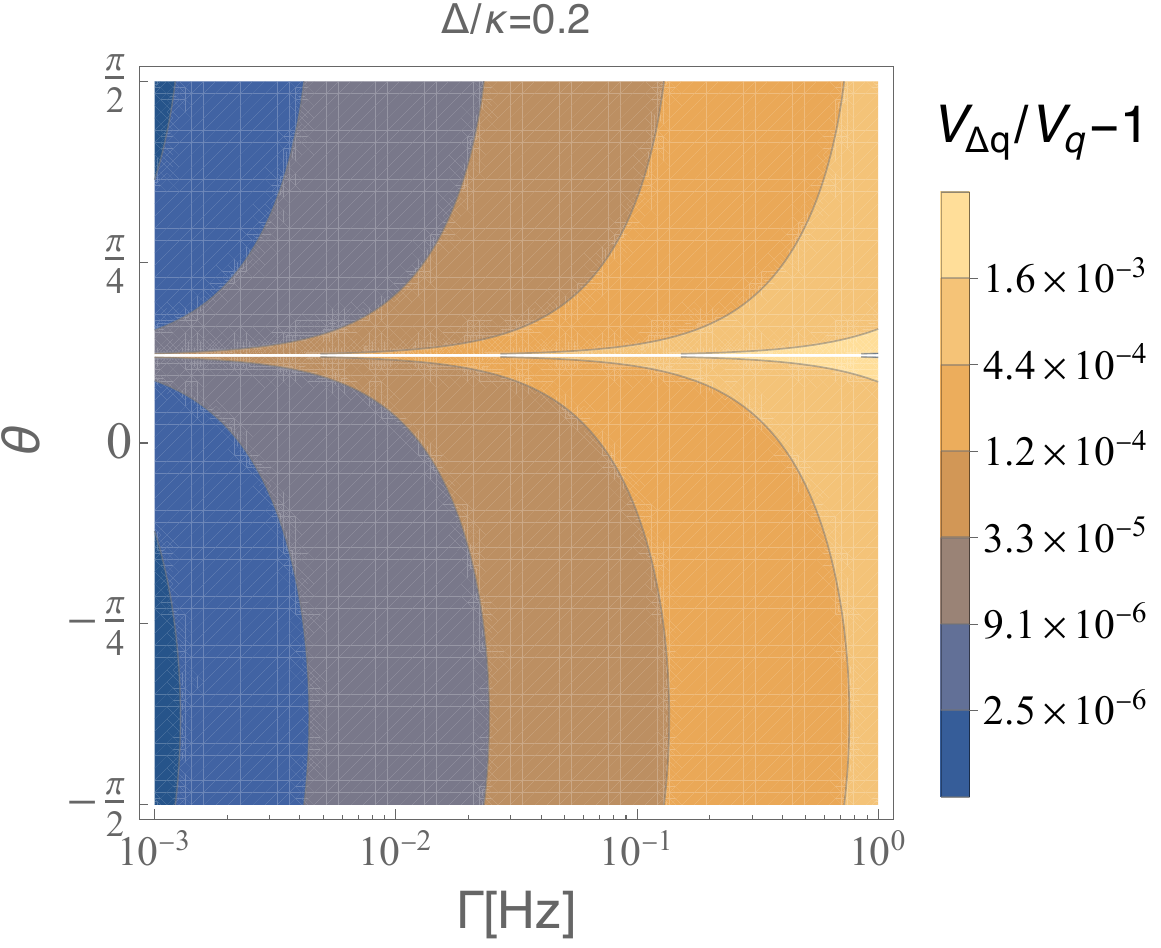}
\hspace{0.5cm}
\includegraphics[width=86mm]{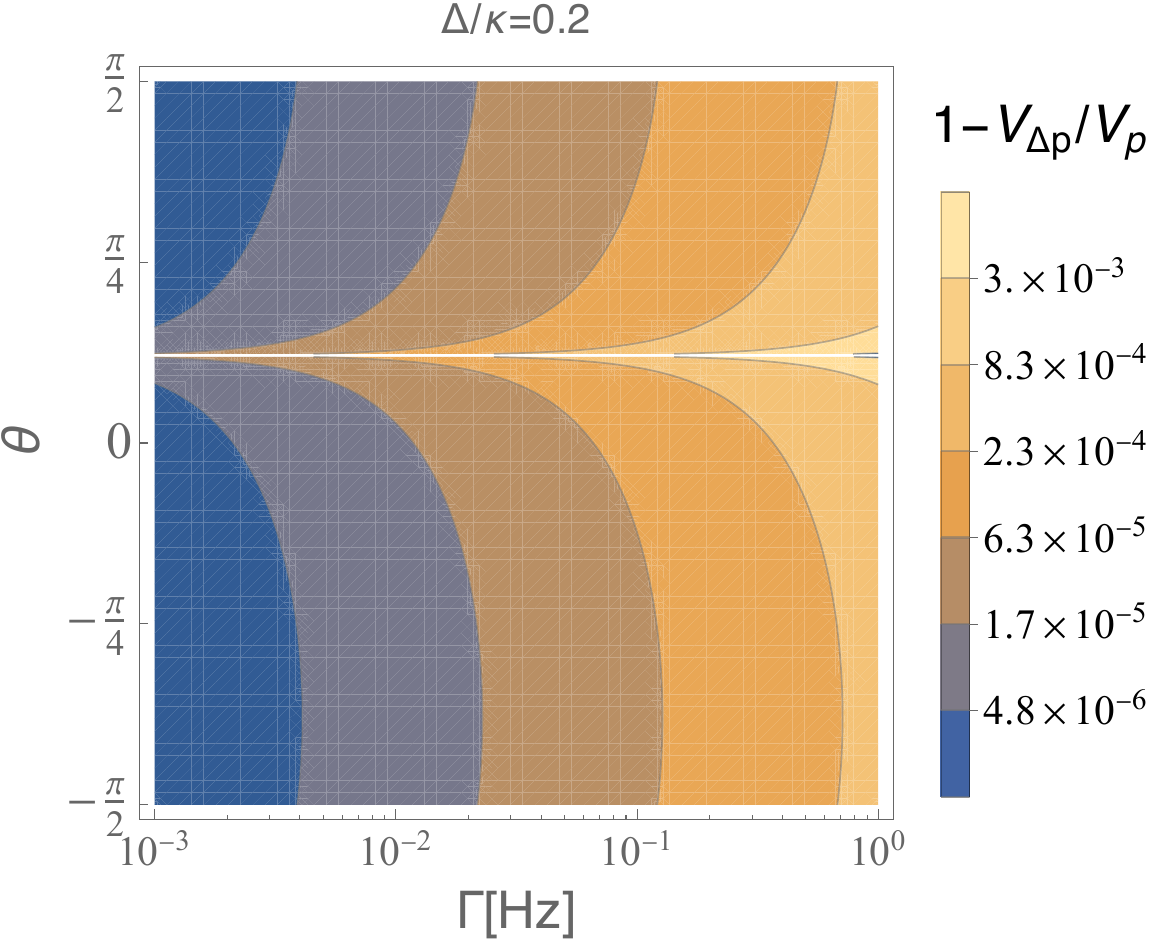}
\caption{These figures show the ratio between the variances obtained from estimation without access to the true system values and those from causal estimation, plotted on the plane spanned by the homodyne angle $\theta$ and the mechanical dissipation rate $\Gamma[\text{Hz}]$. 
The left panel corresponds to the position variance, and the right panel corresponds to the momentum variance. Values close to unity indicate that the variance obtained without access to true values closely matches the causal variance, implying that the reconstruction bias is small. 
Throughout the plots, we fix $\Delta/\kappa = 0.2$, $m=1[\text{mg}]$, $\Omega/2\pi=1[\text{Hz}]$, $\kappa/2\pi=10^8$[Hz] and $P_{\text{in}} = 10^{-5}\text{W}$.
The regions in which the estimation becomes extremely poor correspond to the homodyne angle 
$\theta = \arctan{(2\Delta/\kappa)}$, at which the measurement rate vanishes.
In these plots, the reconstruction bias for both the position and momentum variances remains sufficiently small.}
\label{V_theta_Gamma}
\end{center}
\end{figure} 

\begin{figure}[t]
\begin{center}
\includegraphics[width=86mm]{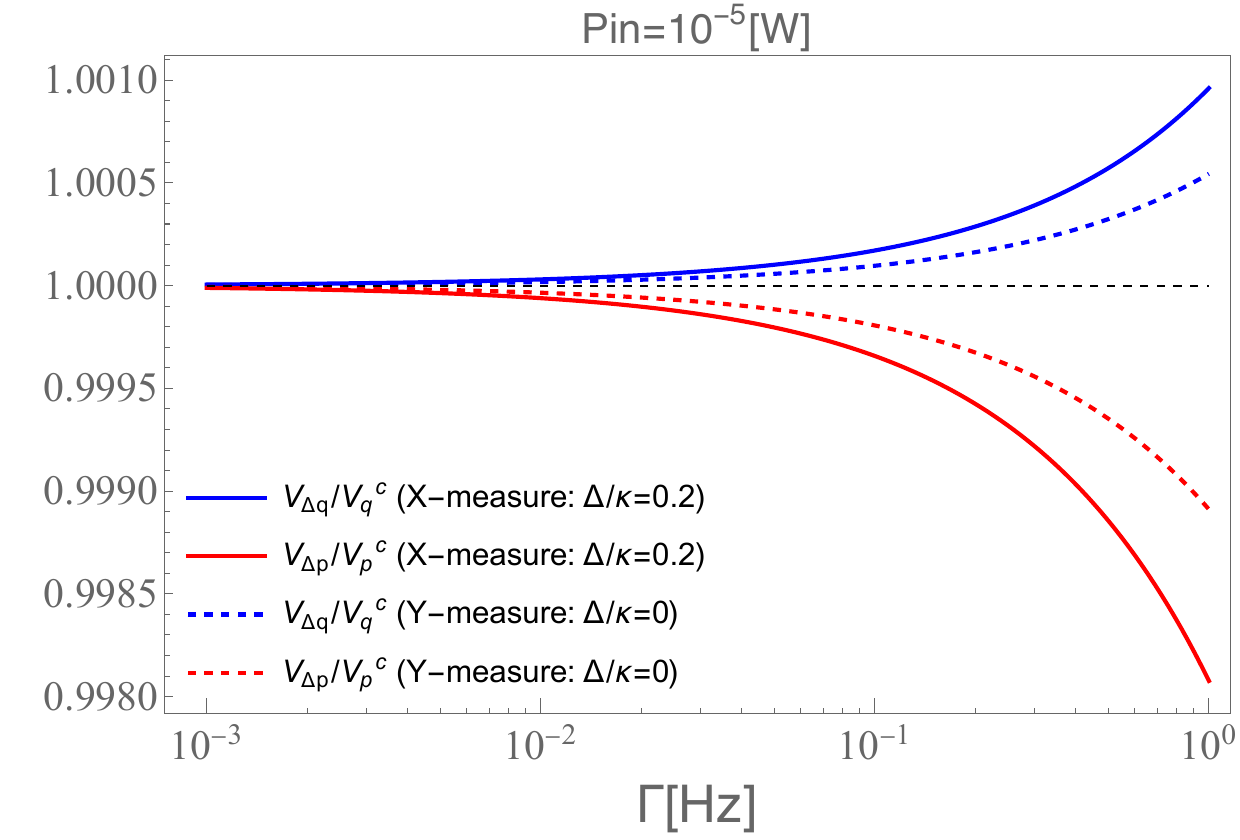}
\hspace{0.5cm}
\includegraphics[width=86mm]{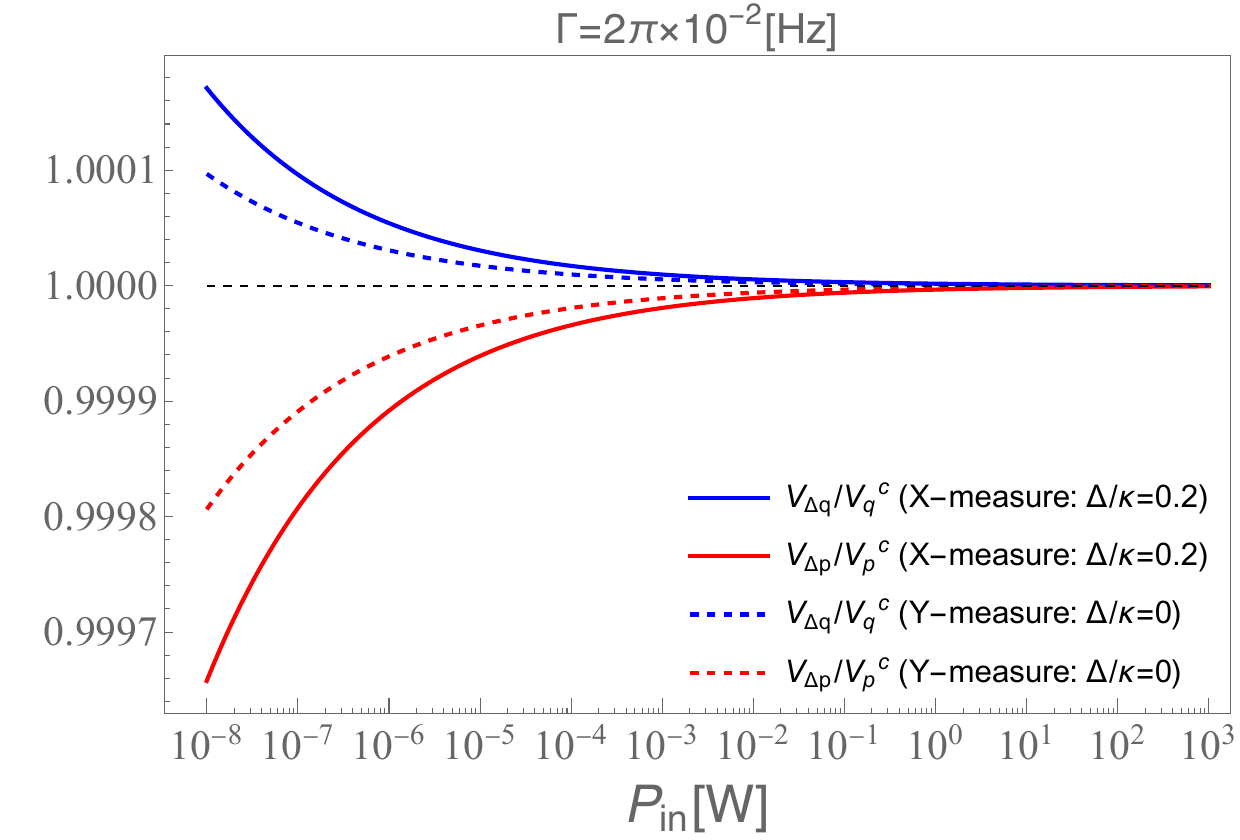}
\caption{These figures show the ratio of the conditional variances reconstructed via measurement-based estimation to those obtained from causal estimation, plotted as a function of~(left) the laser power $P_{\text{in}}[\mathrm{W}]$, and (right) the mechanical damping rate $\Gamma$[Hz]). The blue and red curves represent the ratios for position and momentum, respectively. Solid curves correspond to amplitude (X) measurement $(\theta = 0)$, while dashed curves correspond to phase (Y) measurement $(\theta = \pi/2)$.
For the $Y$-quadrature measurement, we set $\Delta/\kappa=0$, whereas for the $X$-quadrature measurement we use $\Delta/\kappa=0.2$.
For both panels, the common parameters were fixed as $m=1$[mg] and $\Omega/2\pi=1$[Hz]. In the left panel, $P_\text{in}=10^{-5}$[W] was used, whereas in the right panel, $\Gamma/2\pi=\times10^{-2}$[Hz] was fixed.
}
\label{XY-V-ratio}
\end{center}
\end{figure} 
Next, we discuss how the reconstruction bias $\alpha_\theta$ and $\beta_\theta$--which arise in the estimation without access to the true system values--depend on the optomechanical parameters. 
Fig.~\ref{V_theta_Gamma} shows the deviation of the variance obtained from measurement-based estimation, plotted as a function of the homodyne angle $\theta$ and the mechanical dissipation rate $\Gamma~[\mathrm{Hz}]$.
As shown in Fig.~\ref{V_theta_Gamma}, the estimation fails in the vicinity of the homodyne angle $\theta=\arctan(2\Delta/\kappa)$, even when a sufficiently large damping rate $\Gamma$ is introduced. This failure originates from the behavior of the measurement gain, $c_\theta \simeq \sin(\theta-\alpha)$, which vanishes at this angle. 
Consequently, the measured output quadrature carries no information about the oscillator’s position, rendering the estimation procedure ineffective from the outset.

Notably, this homodyne-angle region lies close to the angle at which strong momentum squeezing occurs, given by $\theta_{\mathrm{opt}}=\alpha-\arctan(2/\xi)/2$ with $\xi=\bar{n}/\omega_m$, as discussed in Ref.~\cite{Fukuzumi2025}. A detailed discussion regarding the behavior of the reconstruction bias at $\theta_{\mathrm{opt}}$ is provided in Sec. IV.B.
Outside this narrow angular region, the reconstruction bias exhibits little dependence on the choice of the homodyne angle.

Fig.~\ref{XY-V-ratio} shows the ratio between the variances reconstructed via measurement-based estimation and the conditional variances obtained from causal estimation.
In the left panel, the horizontal axis represents the mechanical dissipation rate, whereas in the right panel it represents the laser power.
The blue curves correspond to the position variance, and the red curves correspond to the momentum variance.
Solid curves indicate the results for the $X$-quadrature measurement $(\theta=0)$, whereas dashed curves correspond to the $Y$-quadrature measurement $(\theta=\pi/2)$.
To ensure the validity of Eqs.~\eqref{bias-Y} and \eqref{bias-X}, the normalized detuning $\Delta/\kappa$ is set to $0$ for the $Y$-quadrature measurement and to $0.2$ for the $X$-quadrature measurement.

From these figures, one finds that the ratio approaches unity as the mechanical dissipation rate $\Gamma$ decreases and the laser power $P_\text{in}$ increases. This indicates that the variances reconstructed via measurement-based estimation converge to the conditional variances obtained from causal estimation.
As described in Eqs.~\eqref{bias-Y} and \eqref{bias-X}, a decrease in $\Gamma$ leads to an enhancement of the cooperativity and the quality factor. Meanfhile, an increase in $P_\text{in}$ enhances the cooperativity and, through the optical spring effect, increases the effective mechanical frequency $\omega_m$, thereby leading to a larger quality factor.

Moreover, in Fig.~\ref{XY-V-ratio}, under sufficiently large laser power  one has $\xi \simeq \kappa/\Delta = 5$, for which the bias is generally smaller for the $Y$-quadrature measurement than for the $X$-quadrature measurement.
However, the red dashed curve corresponding to the $Y$-quadrature measurement indicates that, even in the high-power regime, the measurement-based estimation does not perfectly coincide with the causal estimation.
This behavior follows from Eq.~\eqref{bias-Y}, where the bias for the $Y$-quadrature measurement scales as $\beta_{\pi/2} \propto 1/Q$, leaving a residual contribution proportional to the inverse mechanical the quality factor.
Since this term does not vanish with increasing laser power, the two estimations do not fully coincide unless the quality factor itself is increased.

Assuming ideal gas damping with atomic mass $m_{\rm atom}$ and pressure $P$, the dissipation rate for a cylindrical mirror with radius $R$ and height $h$ is given by $\Gamma =(PR^2\sqrt{8\pi m_{\rm env}}/m\sqrt{k_BT})(1+(h/2R)+\pi/4)$ as derived in Refs.~\cite{Cavalleri2010,Matsumoto2025}.
For an environment dominated by hydrogen atoms, this expression reduces to
\begin{align}
    \Gamma\simeq7\times10^{-6}~{\rm Hz}~\left(P/10^{-3}~{\rm Pa}\right)\left(T/300~{\rm K}\right)^{-1/2}\left(1~{\rm mg}/m\right)^{1/3}\left(20~{\rm g}~{\rm cm}^{-3}/\rho\right)^{2/3}.
\end{align}
Thus, even at room temperature, achieving a vacuum level of $P<1~\mathrm{Pa}$ is feasible, and under such conditions the reconstruction bias becomes essentially negligible.
Even under atmospheric pressure, $P\simeq10^{5}~\mathrm{Pa}$, the corresponding damping rate is only $\Gamma\simeq2~\mathrm{Hz}$, for which the reconstruction bias is still expected to be negligible.
Therefore, when aiming to realize quantum states of a mechanical oscillator—particularly states close to the ground state, as discussed later in this work—we conclude that the reconstruction bias arising from not using the true system values has a negligible impact.

Therefore, in experiments aimed at preparing or realizing quantum states of a mechanical oscillator, even at room temperature, the reconstruction bias is not expected to pose a significant limitation, provided that the environmental damping rate is moderately reduced.

\section{Applications to Quantum Phenomena in Macroscopic Objects}
\subsection{Quantum Entanglement between Macroscopic Objects Using a Power-Recycled Fabry--Perot Interferometer}
In this section, we investigate the generation of quantum entanglement between macroscopic objects using a power-recycled Fabry--Perot cavity.
The fundamental configuration and parameters were proposed in Ref.~\cite{Miki2023}. This system demonstrates that quantum entanglement between milligram-scale oscillators can be generated under realistic experimental conditions by employing continuous measurement and feedback control.

Since the single-mode Hamiltonian is identical to that in the present study and a steady-state causal Kalman filter is employed, the analysis relies on the same covariance matrix as that obtained from the causal Wiener filter investigated here. Therefore, the Wiener-filtering method developed in this paper--which does not rely on the true system values--is directly applicable to their setup. 
The differences between the single-mode system in Ref.~\cite{Miki2023} and the present work arise from the inclusion of feedback control and  structural damping due to suspension wires. Following their feedback scheme, the effective mechanical dissipation rate under feedback is modified from $\Gamma$ to $\gamma_m$ (with $\gamma_m \gg \Gamma$). Consequently, the temperature associated with thermal noise is replaced by an effective temperature $T_\text{eff} = \Gamma T / \gamma_m$. Furthermore, taking into account the structural damping induced by the suspension wires, the effective thermal occupation number is given by $n_{\text{th}} = k_B T_\text{eff} \Omega/(\hbar \omega_m^2)$\cite{Miki2023,Matsumoto}.

Furthermore, two oscillators, denoted as $(q_1, q_2)$, are prepared and positioned as the end mirrors of a Fabry-Perot cavity. The common $(+)$ and the differential $(-)$ modes are defined for both the mechanical motion and the cavity light. Specifically, the corresponding mode operators are given by $q_\pm=(q_1\pm q_2)/\sqrt{2}$, $p_\pm=(p_1\pm p_2)/\sqrt{2}$, and $a_\pm=(a_1\pm a_2)/\sqrt{2}$. Here, the power-recycling mirror modifies the optical cavity decay rate of the common mode, thereby introducing an asymmetry between the modes. This asymmetry can be characterized by introducing a factor $\zeta$ such that $\kappa_+=\kappa_-/\zeta$. It is precisely this mode asymmetry that enables the generation of quantum entanglement.

By appropriately incorporating the parameter modifications induced by this asymmetry, we obtain the relative-estimate covariance matrix of each mode $\displaystyle{V^\pm=\Bigg(\begin{array}{cc}
V^\pm_{\Delta q} & V^\pm_{q p} \\
V^\pm_{q p} & V^\pm_{\Delta p}\end{array}\Bigg)}$, whose components are given by 
\begin{align}
     V_{\Delta q}^\pm&
     =\frac{\gamma_\theta^\pm-\gamma_m}{\lambda_\theta^\pm}+\alpha_\theta^\pm, \qquad   V_{\Delta p}^\pm=
     \frac{\gamma_\theta^\pm-\gamma_m}{\lambda_\theta^\pm}\Bigg[\Bigg(\frac{\omega_\theta^\pm}{\omega_m^\pm}\Bigg)^2-\frac{\gamma_m(\gamma_\theta^\pm-\gamma_m)}{2(\omega_m^\pm)^2}\Bigg]+\beta_\theta^\pm, \qquad
     V_{qp}^{\pm}=\frac{(\gamma_\theta^\pm-\Gamma)^2}{2\lambda_\theta^\pm\omega_m^\pm}.
\end{align}
In this matrix, the causal covariance $V_{qp}^\pm$ is placed in the off-diagonal entries rather than the variational quantity $V_{\Delta q\Delta p}^\pm$. 
Here, our objective is to compute the causal conditional covariance matrix solely from the filter function and the output optical spectrum $S^\pm_{II}(\omega)$, without directly referring to the true system operators  $(\hat{q}, \hat{p})$.
If $V_{\Delta q \Delta p}^\pm$ were placed in the off-diagonal entries, the resulting matrix would fail to reproduce the causal conditional covariance matrix.
The conditional covariance under causal estimation, $V_{qp}$, can be obtained by integrating over the entire frequency domain the real part of the difference between the unconditional covariance and the causal conditional covariance. However, because the unconditional covariance is an odd function of frequency, its integral vanishes and therefore does not contribute to the conditional covariance. Consequently, the causal conditional covariance can be evaluated solely from the output optical spectrum $S_{II}^\pm(\omega)$ and the filter functions. 
\begin{align}
    V^\pm_{qp}:=\int^\infty_{-\infty}\frac{d\omega}{2\pi}\text{Re}\Big[ S^\pm_{qp}(\omega)-\overrightarrow{H}^\pm_q(\omega)\overrightarrow{H}^\pm_p(\omega)^*S_{II}^\pm(\omega)\Big]=-\int^\infty_{-\infty}\frac{d\omega}{2\pi}\text{Re}\Big[\overrightarrow{H}^\pm_q(\omega)\overrightarrow{H}^\pm_p(\omega)^*S_{II}^\pm(\omega)\Big].
\end{align}
The off-diagonal elements of the covariance matrix constructed solely from the estimation quantities are defined by $V_{qp}^\pm$. In this case, provided that the variance estimation is sufficiently accurate, $V^\pm$ faithfully reproduces the causal covariance matrix.
\begin{figure}[t]
\begin{center}
\includegraphics[width=130mm]{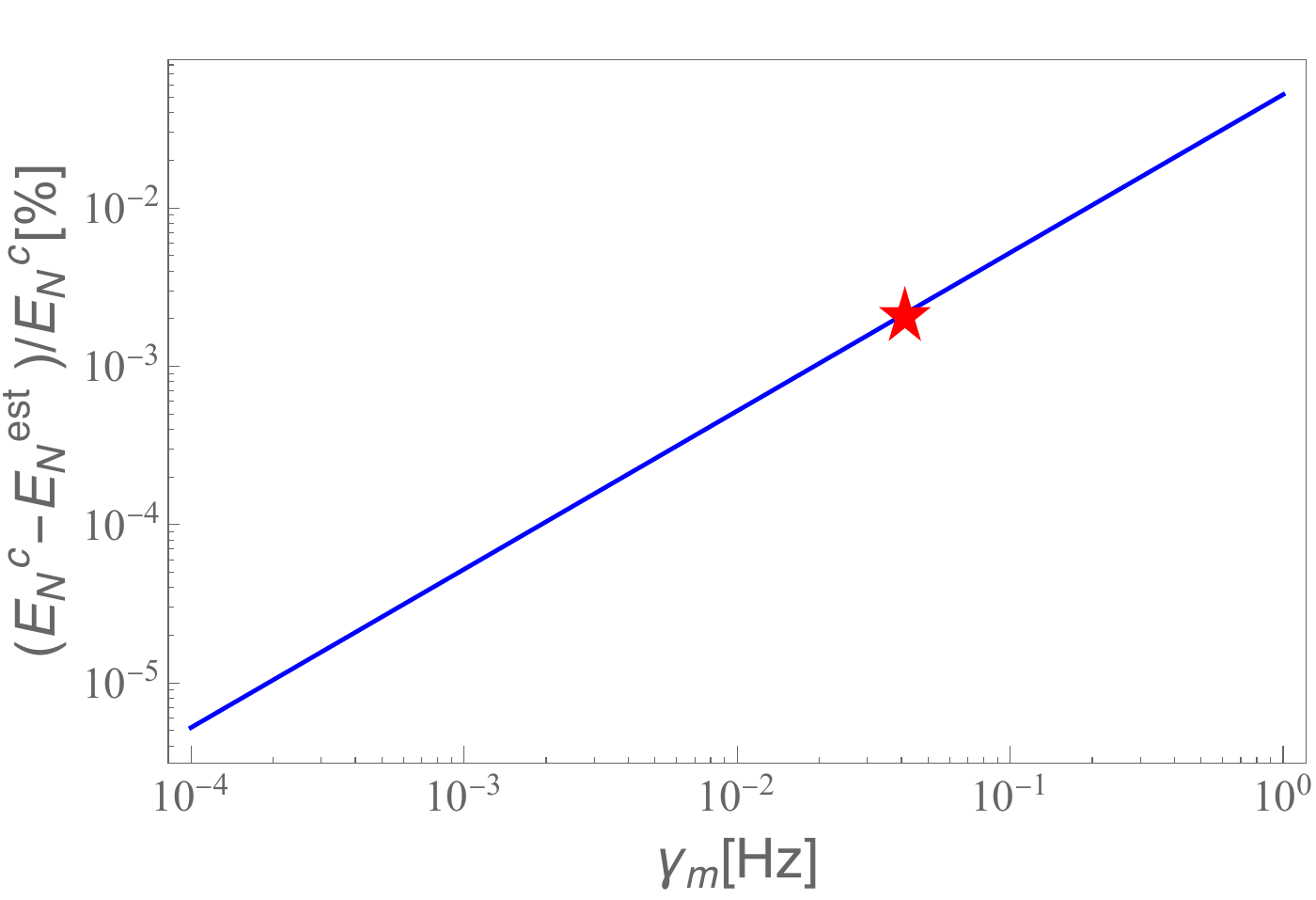} 
\caption{This plot shows the ratio of the logarithmic negativity obtained from measurement-based estimation to that from causal estimation, plotted as a function of the effective mechanical dissipation rate $\gamma_m$. The red star marks the value of $\gamma_m$ used in the previous study \cite{Miki2023}.
Within the parameter range shown, the system remains entangled throughout.
All other parameters are taken from Ref.~\cite{Miki2023}. Representative 
values are $\Omega/2\pi=2.2[\text{Hz}]$, $m=0.92[\text{g}]$, $\Delta/\kappa=0.2$, and $\kappa_-/2\pi=1.64\times10^6[\text{Hz}]$.
The homodyne angle is set to $\theta = 0$, corresponding to an $X$-measurement.
}
\label{elemag_EN}
\end{center}
\end{figure} 
In measurement-based estimation, the two mechanical oscillators are characterized by a $4\times4$ covariance matrix $\mathcal{V}=\Bigg(\begin{array}{cc}
V_{11}&V_{12}\\
        V_{12}&V_{22}
    \end{array}\Bigg)=S\Bigg(\begin{array}{cc}
       V^+&0
        \label{V}
        \\
        0&V^-
    \end{array}\Bigg)S^\text{T}$
that is constructed without reference to the true system values.
The matrix $S$ is defined as
$\displaystyle{S:=\frac{1}{\sqrt{2}}
    \left(\begin{array}{cccc}
        1&0&1&0\\
        0&1&0&1\\
        1&0&-1&0\\
        0&1&0&-1
    \end{array}\right)}$
and represents the beam splitting operation. 
In causal estimation, the conditional covariance matrix for mirrors A and B is obtained by replacing the elements $V_{\Delta q}^{\pm}$ and $V_{\Delta p}^{\pm}$ in $\mathcal{V}$ with the corresponding causal conditional variances, $V_{q}^{\mathrm{c}\pm}$ and $V_{p}^{\mathrm{c}\pm}$, respectively.
The degree of quantum entanglement is quantified by the logarithmic negativity, 
\begin{align}
    E_N&=-\frac{1}{2}\log_2\Bigg(\frac{\Sigma-\sqrt{\Sigma^2-4\det \mathcal{V}}}{2}\Bigg),\qquad
    \Sigma=\det V_{11}+\det V_{22}-2\det V_{12},
\end{align}
which is computed from the covariance matrix.
For a two-mode Gaussian state, entanglement exists if and only if $E_N>0$~\cite{Giedke2001}. 
For the present system in the steady state, the optimization of the logarithmic negativity and the entanglement conditions evaluated using causal estimation have been discussed in Ref.~\cite{Miki2023}. 

Fig.~\ref{elemag_EN} shows the ratio of the logarithmic negativity obtained from causal estimation ($E_N^c$) to that obtained from measurement-based estimation ($E_N^\text{est}$) as a function of the effective mechanical dissipation rate $\gamma_m$ under feedback control. The red star markers indicate the value of $\gamma_m$ considered in the previous study \cite{Miki2023}. All parameters, except for the mechanical dissipation rate, are consistent with those in the previous study.

As shown in Fig.~\ref{elemag_EN}, in the regime of small $\gamma_m$, the difference between the two estimation methods is extremely small. Moreover, the deviation between them increases approximately linearly with $\gamma_m$. This behavior reflects the fact that, as discussed in the previous section, the reconstruction biases $\alpha_\pm$ and $\beta_\pm$ approach zero as the dissipation rate decreases.

Nevertheless, even in the regime of large $\gamma_m$, the deviation from causal estimation remains very small, below $0.1\%$. In particular, the red star marker corresponds to the optimized parameter in the previous study, $\gamma_m/2\pi=6.9\times10^{-3}[\text{Hz}]$, indicating that, within the relevant parameter regime, the deviation from causal estimation can be safely neglected. Therefore, for tabletop experiments aiming at generating quantum entanglement between macroscopic objects, we conclude that the reconstruction bias inherent in measurement-based estimation does not significantly affect the evaluation of entanglement.

Finally, it should be emphasized that $\gamma_m$ cannot be increased arbitrarily.
The causal estimation schemes employed in this work and in Refs.~\cite{Miki2023,Miki2024b,Matsumoto} do not incorporate optimal estimation in the presence of non-Markovian noise associated with feedback control. Accordingly, the formulation based on $\gamma_m$ is valid only in parameter regimes where such feedback-induced noise can be safely neglected. Extending the Wiener and Kalman filtering frameworks to incorporate non-Markovian feedback noise, and developing corresponding measurement-based estimation schemes, constitute important directions for future research.

\subsection{Generation of Significant Momentum Squeezing via Homodyne Measurement}

In this subsection, we apply the measurement-based estimation developed in this work to the conditional states of the same optomechanical system studied in Ref.~\cite{Fukuzumi2025}. In that study, it was shown that, although the position is continuously measured, a momentum-squeezed conditional state can be realized by appropriately choosing the homodyne angle.
In Ref.~\cite{Fukuzumi2025}, the optimal homodyne angle for momentum squeezing was identified as $\theta_{\mathrm{opt}}=\alpha-\arctan(2/\xi)/2$. It was shown that, in the vicinity of this angle, momentum squeezing occurs for sufficiently large laser power. Furthermore, when the mechanical dissipation rate $\Gamma$ is sufficiently small, gravity-induced quantum entanglement is enhanced within this momentum-squeezing regime.
However, this optimal angle lies close to the homodyne angle
$\theta=\alpha-\arctan(2/\xi)/2$,
at which the measurement gain satisfies $c_\theta\simeq0$, causing the estimation procedure to break down. As a result, the impact of estimation errors becomes significant. In this work, we therefore examine how such estimation errors affect the evaluation of conditional states within a verification scheme that does not directly rely on unconditional variances.

\begin{figure}[htbp]
\begin{center}
\includegraphics[width=140mm]{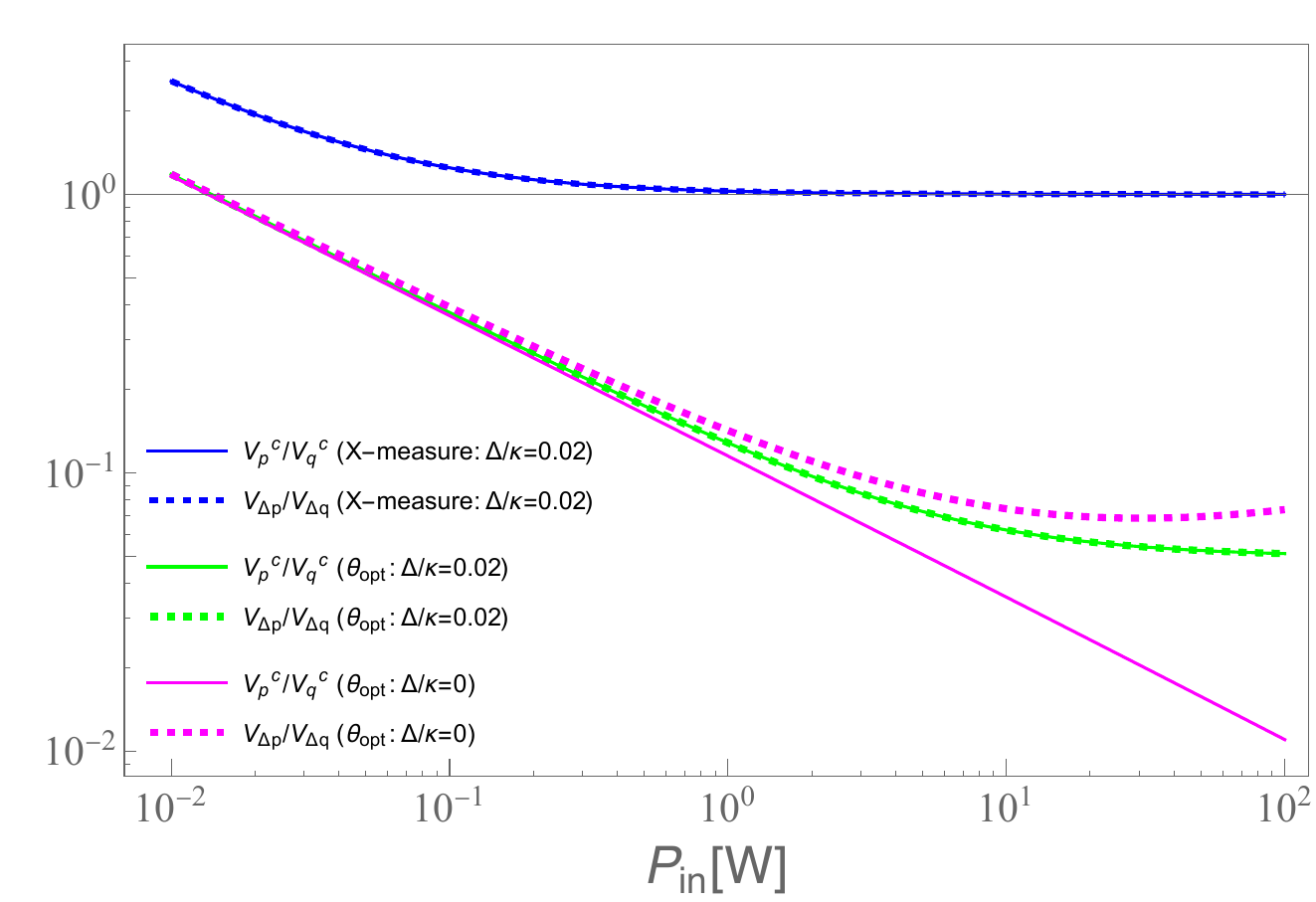}
\caption{
This plot shows the ratio of the momentum variance to the position variance as a function of the input laser power $P_\text{in}$ [W].
The blue curve corresponds to $X$-measurement $(\theta = 0)$ with $\Delta/\kappa = 0.02$.
The green curve represents the case $\Delta/\kappa = 0.02$ with $\theta = \theta_\text{opt}$.
The pink curve corresponds to $\Delta/\kappa = 0$ with $\theta = \theta_\text{opt}$.
Solid curves denote the numerical results obtained from causal estimation, while dashed curves indicate those from measurement-based estimation.
The parameters are fixed at $\Omega/2\pi = 1$ [Hz], $m = 100$ [mg], $\kappa/2\pi = 10^3$ [Hz], and $\Gamma/2\pi = 10^{-2}$ [Hz].}
\label{p-squeez}
\end{center}
\end{figure}

FIG.\ref{p-squeez} shows the ratio of the momentum variance to the position variance plotted as a function of the laser power $P_\text{in}$. Values below unity indicate the presence of momentum squeezing.
The blue curve corresponds to $X$-measurement $(\theta = 0,~\Delta/\kappa = 0.02)$, the green curve to $(\theta = \theta_\text{opt},~\Delta/\kappa = 0.02)$, and the pink curve to $(\theta = \theta_\text{opt},~\Delta/\kappa = 0)$.
Solid curves represent the causal conditional variances, while the dashed curves denote those obtained from measurement-based estimation.
The parameters differ from those used in Ref.~\cite{Fukuzumi2025}. In particular, they are optimized so that momentum squeezing occurs even for a relatively large mechanical dissipation rate of $\Gamma/2\pi = 10^{-2}$ [Hz].
It should also be emphasized that, under this optimized parameter set, momentum squeezing is realized in the sense $V_p/V_q < 1$, although the absolute conditional variance does not satisfy $V_p^{\mathrm{c}} < 1$.

As shown in Fig.~\ref{p-squeez}, for $X$-quadrature measurement the position and momentum variances converge to the same value as the laser power increases. In contrast, when the optimal homodyne angle $\theta_\text{opt}$ is employed, strong momentum squeezing is induced.
Moreover, when the detuning is nonzero, the difference between the solid and dashed curves is negligible, indicating that the reconstruction bias can be safely ignored in this regime.
In contrast, for zero detuning with $\theta = \theta_\text{opt}$, the variance ratio obtained from causal estimation increasingly deviates from that obtained via measurement-based estimation as $P_\text{in}$ grows. 
This behavior is opposite to the tendency discussed in the previous sections.

In the case $\theta=\theta_{\text{opt}}$ and $\Delta=0$, since the detuning vanishes, the optical spring effect is absent and $\omega_m = \Omega$, one has the approximations $\Lambda_{\theta_{\text{opt}}}\simeq -\Omega$ and $\gamma_{\theta_{\text{opt}}}^2\simeq 2\omega_{\theta_{\text{opt}}}^2$, which imply $G\simeq \Omega^4$.
Under these conditions, when the mechanical dissipation rate is sufficiently small $(\Gamma \ll \Omega)$ and the laser power is sufficiently large, the reconstructed bias can be approximated as
\begin{align}
    \alpha_{\theta_{\text{opt}}}\simeq\frac{2\xi}{Q},\qquad \beta_{\theta_{\text{opt}}}\simeq \frac{\xi}{Q}.\label{bias-opt-angle}
\end{align}
Because the detuning is zero $\Delta=0$, the optical spring effect does not enhance the effective mechanical frequency, and increasing the laser power does not improve the quality factor $Q$. Furthermore, in this regime 
$\xi$ does not approach a constant value with increasing laser power but instead continues to increase. Consequently, the bias in Eq.~\eqref{bias-opt-angle} increases with $P_{\rm in}$.
At $P_\text{in} = 1$ [W], the variance ratio obtained from causal estimation differs from that obtained via measurement-based estimation by as much as $23\%$. This feature is unique to the case where $\theta=\theta_\text{opt},~\Delta = 0$.

Therefore, under this parameter set, a nonzero detuning is preferable for verifying momentum squeezing, as it stabilizes the estimation and suppresses the growth of reconstruction bias.
More generally, if a lower-dissipation environment can be realized such that a mechanical quality factor is substantially improved, the discrepancy arising from these biases would be further reduced.
In particular, if the extremely small dissipation rate required for realizing gravity-induced quantum entanglement, $\Gamma \sim 10^{-18}\mathrm{Hz}$, can be achieved, the resulting increase in the quality factor would render the reconstruction bias practically negligible.

\section{Conclusion}
In this work, we investigated measurement-based estimation of the quantum state of a mechanical oscillator in a cavity optomechanical system and assessed its validity within a consistent theoretical framework based on the quantum Langevin equations, including both thermal and radiation-pressure noise.
For a detuned cavity under homodyne detection, we formulated causal and anti-causal Wiener filters and introduced relative-estimate operators $(\Delta q,\Delta p)$ defined solely in terms of the measured output field and the corresponding filter functions. By analytically evaluating their variances and performing numerical comparisons, we demonstrated that these quantities closely reproduce the causal conditional variances over a broad parameter regime. In particular, by appropriately combining past (causal) and future (anti-causal) estimations, the conditional variances can be reconstructed with high accuracy without relying on the unconditional variances.

In the strong laser-driving regime, we derived simple approximate expressions for the reconstruction bias. For phase ($Y$) measurement, the position reconstruction bias scales inversely with the cooperativity, while the momentum reconstruction bias is governed by the inverse mechanical quality factor. For amplitude ($X$) measurement, finite detuning is required to achieve measurement sensitivity, and the reconstruction bias is determined by the ratio of the normalized radiation-pressure noise to the mechanical quality factor. Importantly, the required low-dissipation and low-noise conditions remain experimentally accessible within current optomechanical platforms.

We applied this framework to the verification of macroscopic quantum entanglement, focusing on the electromagnetically mediated entanglement proposed in Ref.~\cite{Miki2023}. By reconstructing the covariance matrix using the present measurement-based estimation scheme, we showed that the logarithmic negativity is accurately reproduced and that the reconstruction bias has a negligible impact in the experimentally relevant parameter regime.
We further examined the conditional momentum-squeezed states discussed in Ref.~\cite{Fukuzumi2025}. While momentum squeezing can be realized by selecting the optimal homodyne angle $\theta_\text{opt}$, we analytically showed that, in the absence of detuning, increasing the laser power enhances the reconstruction bias. This contrasts with the behavior observed in standard $X$ and $Y$ measurements and highlights the critical role of detuning in ensuring robust state verification.

As an important future direction, it will be necessary to explicitly incorporate feedback terms into the equations of motion and extend the present estimation framework to include feedback-induced (potentially non-Markovian) noise. Since feedback is primarily employed to increase the effective dissipation rate and thereby shorten the measurement time, such an extension is particularly relevant for long-time experiments aimed at observing gravity-induced quantum entanglement \cite{Miki2024b,Matsumoto2025,Hatakeyama2025}. Achieving high-precision measurement-based state reconstruction under realistic feedback conditions will therefore be essential for advancing macroscopic quantum tests.

\section{Acknowledgement}
We thank Prof.~Yanbei Chen for insightful discussions.
This work is supported by JSPS KAKENHI Grant No.~JP23H01175, and D.M. is supported by the JSPS Overseas Research Fellowships.

\section*{Appendix A}
In this Appendix, we briefly review and derive the causal/anti-causal quantum Wiener filter.
First, in Fourier domain, the spectrum density of the output optical quadrature is given by
\begin{align}
    S_{II}(\omega)=S_{II}^\text{C}(\omega)S_{II}^\text{AC}(\omega),
\end{align}
where $S_{II}^{\mathrm{C}}(\omega)$ denotes the causal part of $S_{II}(\omega)$, while $S_{II}^{\mathrm{AC}}(\omega)$ denotes its anti-causal part. These components satisfy $(S_{II}^{\mathrm{C}}(\omega))^* = S_{II}^{\mathrm{AC}}(\omega)$.
We define the causal operator $\tilde{I}_\theta(\omega)$, and we can obtain the normalized spectrum of output quadrature,
\begin{align}
    \tilde{I}_\theta(\omega)=\frac{\hat{I}_\theta(\omega)}{S_{II}^\text{C}(\omega)}, \qquad S_{\tilde{I}\tilde{I}}(\omega)=1.
\end{align}
Note also that the correlation of this normalized output quadrature in the time domain is given by $\langle \tilde{I}(t)\tilde{I}(t-s)\rangle = \delta(s)$.
The cross spectrum of the oscillator's position and output quadrature, 
\begin{align}
    S_{q\tilde{I}}(\omega)=\frac{S_{qI}(\omega)}{S_{II}^\text{AC}(\omega)}.
\end{align}

Next, we consider the causal quantum wiener filter for position.
In time domain, the causal estimate value is defined as $\displaystyle{\overrightarrow{q}(t)=\int^t_{-\infty}dt'\overrightarrow{h}'(t-t')\tilde{I}_\theta(t')}$, where $\overrightarrow{h}'(t-t')$ is the filter that allows optimal estimation for the normalized output quadrature.
To minimize the mean-square value between the system operator $\hat{q}$ and the estimator $\overrightarrow{q}$, one simply requires that the variation of the functional with respect to $h$ vanishes.
\begin{align}
    0&=\delta_{h'(s-s')} \Big\langle{\Big(\hat{q}(t)-\overrightarrow{q}(t)\Big)\Big(\hat{q}(s)-\overrightarrow{q}(s)\Big)}\Big\rangle\nonumber\\
    &=\delta_{h'(s-s')}\Big(\Big \langle \hat{q}(t)\hat{q}(s) \Big\rangle - \Big\langle \int^t_{-\infty}dt'\overrightarrow{h}'(t-t')\tilde{I}_\theta(t')\hat{q}(s) \Big\rangle\nonumber\\
    &\quad-\Big\langle \hat{q}(t)\int^s_{-\infty}ds'\overrightarrow{h}'(s-s')\tilde{I}_\theta(s')\Big\rangle + \Big\langle \int^t_{-\infty}dt'\overrightarrow{h}'(t-t')\tilde{I}_\theta(t')\int^s_{-\infty}ds'\overrightarrow{h}'(s-s')\tilde{I}_\theta(s')\Big\rangle \Big)\nonumber\\
    &=-\Big\langle\Big(\hat{q}(t)-\overrightarrow{q}(t)\Big)\int^s_{-\infty}ds'\delta\overrightarrow{h}'(s-s')\tilde{I}_\theta(s')\Big\rangle\nonumber\\
    &=-\int^t_{-\infty}ds'\delta\overrightarrow{h}'(s-s')\Big\langle \Big( \hat{q}(t)-\int^t_{-\infty}dt'\overrightarrow{h}'(t-t')\tilde{I}_\theta(t')\Big)\tilde{I}_\theta(s)\Big\rangle.\label{filter-condition}
\end{align}
Thus, we obtain the condition of the causal optimal filter, $\displaystyle{\Big\langle \Big( \hat{q}(t)-\int^t_{-\infty}dt'\overrightarrow{h}'(t-t')\tilde{I}_\theta(t')\Big)\tilde{I}_\theta(s)\Big\rangle=0}$. 
This condition indicates that there are no-correlation between the residual uncertainty after optimal estimation and the measurement data.
By changing the integration variables to $\lambda = t - t'$ and $s=t-\tau$, we can rewrite the above condition.
\begin{align}
    \Big\langle \hat{q}(t) \tilde{I}(t-\tau)\Big\rangle &=  \int^\infty_0d\lambda\overrightarrow{h}'(\lambda)\Big\langle\tilde{I}_\theta(t-\lambda)\tilde{I}_\theta(t-\tau) \Big\rangle\nonumber\\
    &=\int^\infty_0d\lambda\overrightarrow{h}'(\lambda)\delta(\tau-\lambda).
\end{align}
So we obtain the solution of the optimal filter, which reflects the fact that it is causal, taking nonzero values only for $\tau \geq 0$.
\begin{align}
    \overrightarrow{h}'(\tau) = 
\left\{
\begin{array}{ll}
\langle \hat{q}(t) \tilde{I}(t-\tau)\rangle & (\tau \geq 0)\\
0 & (\tau < 0)
\end{array}
\right.
\end{align}
Next, we consider the Wiener filter corresponding to the Fourier transform of this causal filter. Since $\overrightarrow{h}'(\tau)$ has support only for $\tau \geq 0$, its Fourier transform can be defined using a one-sided integral.
\begin{align}
    \overrightarrow{H}'(\omega)&=\int^\infty_0d\tau \overrightarrow{h}'(\tau)e^{i\omega\tau}=\int^\infty_0d\tau \Big\langle \hat{q}(t) \tilde{I}(t-\tau)\Big\rangle e^{i\omega\tau}.
\end{align}
On the other hand, the cross-spectrum between $\hat{q}(\omega)$ and $\tilde{I}_\theta(\omega)$, which corresponds to the Fourier transform of $\langle \hat{q}(t)\tilde{I}(t-\tau)\rangle$, is defined over the entire time domain. 
Therefore, if we explicitly separate the integration range, we obtain
\begin{align}
    S_{q\tilde{I}}(\omega)&=\int^\infty_{-\infty}d\tau \Big\langle \hat{q}(t) \tilde{I}(t-\tau)\Big\rangle e^{i\omega\tau}\nonumber\\
    &=\Bigg(\int^\infty_0d\tau+\int^0_{-\infty}d\tau\Bigg)\Big\langle \hat{q}(t) \tilde{I}(t-\tau)\Big\rangle e^{i\omega\tau}\nonumber\\
    &=\Big[S_{q\tilde{I}}(\omega)\Big]_++\Big[S_{q\tilde{I}}(\omega)\Big]_-.
\end{align}
Now, we assume $\tau\geq0$, the integration range $0\leq\tau<\infty$ represents "causal". 
Here, $[\cdots]_{+/-}$ explicitly denotes the causal/anti-causal function. Consequently, the Wiener filter for estimating the system using the normalized output quadrature is given by
\begin{align}
    \overrightarrow{H}'(\omega)=\Big[S_{q\tilde{I}}(\omega)\Big]_+=\Bigg[\frac{S_{q\hat{I}}(\omega)}{S_{II}^\text{AC}(\omega)}\Bigg]_+.
\end{align}
Since the estimation operator is given by $\overrightarrow{q}(\omega)=\overrightarrow{H}'(\omega)\tilde{I}_\theta(\omega)=\overrightarrow{H}(\omega)\hat{I}_\theta(\omega)$, restoring the original (unnormalized) form of $\tilde{I}_\theta(\omega)$ shows that the Wiener filter is 
\begin{align}
    \overrightarrow{H}(\omega)=
    \frac{1}{S_{II}^\text{C}(\omega)}\Bigg[\frac{S_{q\hat{I}}(\omega)}{S_{II}^\text{AC}(\omega)}\Bigg]_+.
\end{align}
\vskip\baselineskip

On the other hand, one can perform optimal estimation using information obtained after a given time $t$, which we refer to as anti-causal filtering. The basic procedure of the calculation is the same as that for the causal estimation discussed above.
In this case, the normalized output quadrature should be defined as $\tilde{I}_\theta(\omega)=\hat{I}_\theta(\omega)/S_{II}^{\mathrm{AC}}(\omega)$. The condition of the optimal filter in the time domain is rewritten as 
\begin{align}
    0=\Bigg\langle \Bigg(\hat{q}(t)-\int^\infty_tdt'\overleftarrow{h}'(t-t') \tilde{I}_\theta(t')\Bigg) \tilde{I}_\theta(s)\Bigg\rangle.
\end{align}
Here we assume $\lambda=t-t'\leq0$, since anti-causal filtering requires the condition $t\leq t'<\infty$.
By changing the integration variable to $\lambda$, we obtain 
\begin{align}
    \Big\langle\hat{q}(t)\tilde{I}_\theta(s)\Big\rangle&=\int^\infty_t dt'\overleftarrow{h}'(t-t')\Big\langle\tilde{I}_\theta(t')\tilde{I}_\theta(s)\Big\rangle\nonumber\\
     &=\int^\infty_tdt'\overleftarrow{h}'(t-t')\delta(t'-s)\nonumber\\
     &=\int^0_{-\infty} d\lambda \overleftarrow{h}'(\lambda)\delta(\lambda-\tau).
\end{align}
where we introduce $\tau=t-s$.
Since the integration variable $\lambda\leq0$ is negative, the above expression is a nonzero value only for $\tau\leq0$. Therefore, 
\begin{align}
    \overleftarrow{h}'(\tau) = 
\left\{
\begin{array}{ll}
\langle \hat{q}(t) \tilde{I}(t-\tau)\rangle & (\tau \leq 0)\\
0 & (\tau > 0)
\end{array}
\right.
\end{align}
In the anti-causal regime, since $\tau\leq0$, the filter is nonzero only in the future-time domain.
Since the anti-causal optimal filter in the time domain is non-zero only for $\tau\leq0$, its Fourier transform is written as 
\begin{align}
    \overleftarrow{H}'(\omega)&=\int^0_{-\infty}d\tau\overleftarrow{h}'(\tau)e^{-i\omega\tau}\nonumber\\
    &=\int^0_{-\infty}d\tau\Big\langle\hat{q}(t)\tilde{I}_\theta(t-\tau)\Big\rangle e^{-i\omega\tau}\nonumber\\
    &=\Big[S_{q\tilde{I}}(\omega)^*\Big]_-,
    \label{AC-WF'}
\end{align}
where we use the relation $S_{q\tilde{I}}(-\omega)=S_{q\tilde{I}}(\omega)^*$.
For the anti-causal filter, the time direction is reversed relative to the causal one. Correspondingly, the Fourier kernel is replaced from  $e^{i\omega \tau}$ by $e^{-i\omega\tau}$.
Therefore, Eq.~\eqref{AC-WF'} is defined using the complex conjugate of the cross-spectrum evaluated under the causal Fourier-transform convention.
Finally, we obtain the definition of the anti-causal wiener filter,
\begin{align}
    \overleftarrow{H}(\omega)=
    \frac{1}{S_{II}^\text{AC}(\omega)}\Bigg[\frac{S_{qI}(\omega)^*}{S_{II}^\text{C}(\omega)}\Bigg]_-.
\end{align}

\section*{Appendix. B}
In this appendix, we summarize useful integral formulas that are used to evaluate entire frequency-domain integrals in measurement-based estimation.
\begin{align}
    &\frac{1}{2\pi}\int^\infty_{-\infty}\frac{\omega d\omega}{|F(\omega)|^2F'(\omega)}=\frac{i(\gamma_\theta+\Gamma)}{2\Gamma G},\nonumber\\
    &\frac{1}{2\pi}\int^\infty_{-\infty}\frac{\omega^2 d\omega}{|F(\omega)|^2F'(\omega)}=\frac{\omega^{2}_\theta-\omega^{2}_m}{2\Gamma G},\nonumber\\
    &\frac{1}{2\pi}\int^\infty_{-\infty}\frac{\omega^3 d\omega}{|F(\omega)|^2F'(\omega)}=\frac{-i(\gamma_\theta\omega^{2}_m+\Gamma\omega^{2}_\theta)}{2\Gamma G},\nonumber\\
    &\frac{1}{2\pi}\int^\infty_{-\infty}\frac{\omega^4 d\omega}{|F(\omega)|^2F'(\omega)}=-\frac{\Gamma\gamma_\theta\omega^{2}_m+\Gamma^2\omega^{2}_\theta-\omega^{2}_\theta\omega^{2}_m+\omega^{4}_m}{2\Gamma G},\nonumber
    \end{align}
    and
    \begin{align}
    &\frac{1}{2\pi}\int^\infty_{-\infty}\frac{\omega d\omega}{|F(\omega)|^2(F'(\omega))^2}=\frac{i(\gamma_\theta+\Gamma)(2\omega^{2}_\theta-2\omega^{2}_m+(\gamma_\theta-\Gamma)\Gamma)}{2\Gamma G^2},\nonumber\\
    &\frac{1}{2\pi}\int^\infty_{-\infty}\frac{\omega^2 d\omega}{|F(\omega)|^2(F'(\omega))^2}=\frac{-(\gamma_\theta-\Gamma)\omega^{2}_m-2\omega^{2}_m\omega^{2}_\theta+\omega^{4}_m+\omega^{4}_\theta}{2\Gamma G^2},\nonumber\\
    &\frac{1}{2\pi}\int^\infty_{-\infty}\frac{\omega^3 d\omega}{|F(\omega)|^2(F'(\omega))^2}=-\frac{i(\omega^{2}_\theta-\omega^{2}_m)((\Gamma+2\gamma_\theta)\omega^{2}_m+\Gamma\omega^{2}_\theta)}{2\Gamma G^2},\nonumber\\
    &\frac{1}{2\pi}\int^\infty_{-\infty}\frac{\omega^4 d\omega}{|F(\omega)|^2(F'(\omega))^2}=-\frac{(\omega^{2}_m(\gamma_\theta+\omega_m)+\omega^{2}_\theta(\Gamma-\omega_m))(\omega^{2}_m(\gamma_\theta-\omega_m)+\omega^{2}_\theta(\Gamma+\omega_m))}{2\Gamma G^2},\nonumber
\end{align}
where we defined $G=\gamma_\theta^2\omega_m^2+\Gamma^2\omega_\theta^2+(\omega_\theta^2-\omega_m^2)^2+\gamma_\theta\Gamma(\omega_\theta^2+\omega_m^2)$. 

\bibliography{reference}

@article{Miki2024a,
  title = {Quantum signature of gravity in optomechanical systems with conditional measurement},
  author = {Miki, Daisuke and Matsumura, Akira and Yamamoto, Kazuhiro},
  journal = {Phys. Rev. D},
  volume = {109},
  issue = {6},
  pages = {064090},
  numpages = {12},
  year = {2024},
  month = {Mar},
  publisher = {American Physical Society},
  doi = {10.1103/PhysRevD.109.064090},
  url = {https://link.aps.org/doi/10.1103/PhysRevD.109.064090}
}

@article{Miki2024b,
  title = {Feasible generation of gravity-induced entanglement by using optomechanical systems},
  author = {Miki, Daisuke and Matsumura, Akira and Yamamoto, Kazuhiro},
  journal = {Phys. Rev. D},
  volume = {110},
  issue = {2},
  pages = {024057},
  numpages = {5},
  year = {2024},
  month = {Jul},
  publisher = {American Physical Society},
  doi = {10.1103/PhysRevD.110.024057},
  url = {https://link.aps.org/doi/10.1103/PhysRevD.110.024057}
}

@article{Miki2023,
  title = {Generating quantum entanglement between macroscopic objects with continuous measurement and feedback control},
  author = {Miki, Daisuke and Matsumoto, Nobuyuki and Matsumura, Akira and Shichijo, Tomoya and Sugiyama, Yuuki and Yamamoto, Kazuhiro and Yamamoto, Naoki},
  journal = {Phys. Rev. A},
  volume = {107},
  issue = {3},
  pages = {032410},
  numpages = {15},
  year = {2023},
  month = {Mar},
  publisher = {American Physical Society},
  doi = {10.1103/PhysRevA.107.032410},
  url = {https://link.aps.org/doi/10.1103/PhysRevA.107.032410}
}

@article{Miao2020,
  title = {Quantum correlations of light mediated by gravity},
  author = {Miao, Haixing and Martynov, Denis and Yang, Huan and Datta, Animesh},
  journal = {Phys. Rev. A},
  volume = {101},
  issue = {6},
  pages = {063804},
  numpages = {7},
  year = {2020},
  month = {Jun},
  publisher = {American Physical Society},
  doi = {10.1103/PhysRevA.101.063804},
  url = {https://link.aps.org/doi/10.1103/PhysRevA.101.063804}
}

@article{Datta2021,
doi = {10.1088/2058-9565/ac1adf},
url = {https://doi.org/10.1088/2058-9565/ac1adf},
year = {2021},
month = {aug},
publisher = {IOP Publishing},
volume = {6},
number = {4},
pages = {045014},
author = {Datta, Animesh and Miao, Haixing},
title = {Signatures of the quantum nature of gravity in the differential motion of two masses},
journal = {Quantum Science and Technology}
}

@article{Hatakeyama2025,
    author = "Hatakeyama, Kosei and Miki, Daisuke and Yamamoto, Kazuhiro",
    title = "{Theoretical study of the squeezed-light-enhanced sensitivity to gravity-induced entanglement via finite-time analysis}",
    doi = "10.1103/1mfv-y24t",
    journal = "Phys. Rev. D",
    volume = "113",
    number = "2",
    pages = "024025",
    year = "2026"
}

@misc{fukuzumi2025,
      title={Momentum Squeezed State Realized via Optimal Filtering in Optomechanics: Implications for Gravity-Induced Entanglement}, 
      author={Ryotaro Fukuzumi and Kosei Hatakeyama and Daisuke Miki and Kazuhiro Yamamoto},
      year={2026},
      eprint={2508.14337},
      archivePrefix={arXiv},
      primaryClass={quant-ph},
      url={https://arxiv.org/abs/2508.14337}, 
}

@misc{Matsumoto,
      title={Verification of conditional mechanical squeezing for a mg-scale pendulum near quantum regimes}, 
      author={Jordy G. Santiago-Condori and Naoki Yamamoto and Nobuyuki Matsumoto},
      year={2023},
      eprint={2008.10848},
      archivePrefix={arXiv},
      primaryClass={quant-ph},
      url={https://arxiv.org/abs/2008.10848}, 
}

@article{Gardiner1985,
  title = {Input and output in damped quantum systems: Quantum stochastic differential equations and the master equation},
  author = {Gardiner, C. W. and Collett, M. J.},
  journal = {Phys. Rev. A},
  volume = {31},
  issue = {6},
  pages = {3761--3774},
  numpages = {0},
  year = {1985},
  month = {Jun},
  publisher = {American Physical Society},
  doi = {10.1103/PhysRevA.31.3761},
  url = {https://link.aps.org/doi/10.1103/PhysRevA.31.3761}
}

@misc{Matsumoto2025,
      title={Space-based cm/kg-scale Laser Interferometer for Quantum Gravity}, 
      author={Nobuyuki Matsumoto and Katsuta Sakai and Kosei Hatakeyama and Kiwamu Izumi and Daisuke Miki and Satoshi Iso and Akira Matsumura and Kazuhiro Yamamoto},
      year={2025},
      eprint={2507.12899},
      archivePrefix={arXiv},
      primaryClass={gr-qc},
      url={https://arxiv.org/abs/2507.12899}, 
}

@article{Aspelmeyer2014,
  title = {Cavity optomechanics},
  author = {Aspelmeyer, Markus and Kippenberg, Tobias J. and Marquardt, Florian},
  journal = {Rev. Mod. Phys.},
  volume = {86},
  issue = {4},
  pages = {1391--1452},
  numpages = {62},
  year = {2014},
  month = {Dec},
  publisher = {American Physical Society},
  doi = {10.1103/RevModPhys.86.1391},
  url = {https://link.aps.org/doi/10.1103/RevModPhys.86.1391}
}

@misc{Shichijo,
      title={Quantum state of a suspended mirror coupled to cavity light -- Wiener filter analysis of the pendulum and rotational modes}, 
      author={Tomoya Shichijo and Nobuyuki Matsumoto and Akira Matsumura and Daisuke Miki and Yuuki Sugiyama and Kazuhiro Yamamoto},
      year={2023},
      eprint={2303.04511},
      archivePrefix={arXiv},
      primaryClass={quant-ph},
      url={https://arxiv.org/abs/2303.04511}, 
}

@book{Wiener1949,
  author       = {Norbert Wiener},
  title        = {Extrapolation, Interpolation, and Smoothing of Stationary Time Series: With Engineering Applications},
  publisher    = {The MIT Press},
  year         = {1949},
  doi          = {10.7551/mitpress/2946.001.0001},
  isbn         = {9780262257190},
  url          = {https://doi.org/10.7551/mitpress/2946.001.0001}
}

@article{Chen2013,
    author = "Chen, Yanbei",
    title = "{Macroscopic Quantum Mechanics: Theory and Experimental Concepts of Optomechanics}",
    eprint = "1302.1924",
    archivePrefix = "arXiv",
    primaryClass = "quant-ph",
    reportNumber = "LIGO-P1300014, LIGO-P1300014",
    doi = "10.1088/0953-4075/46/10/104001",
    journal = "J. Phys. B",
    volume = "46",
    pages = "104001",
    year = "2013"
}

@article{Meng2022,
author = {Chao Meng  and George A. Brawley  and Soroush Khademi  and Elizabeth M. Bridge  and James S. Bennett  and Warwick P. Bowen },
title = {Measurement-based preparation of multimode mechanical states},
journal = {Science Advances},
volume = {8},
number = {21},
pages = {eabm7585},
year = {2022},
doi = {10.1126/sciadv.abm7585},
}

@article{Rossi,
  title = {Observing and Verifying the Quantum Trajectory of a Mechanical Resonator},
  author = {Rossi, Massimiliano and Mason, David and Chen, Junxin and Schliesser, Albert},
  journal = {Phys. Rev. Lett.},
  volume = {123},
  issue = {16},
  pages = {163601},
  numpages = {6},
  year = {2019},
  month = {Oct},
  publisher = {American Physical Society},
  doi = {10.1103/PhysRevLett.123.163601},
  url = {https://link.aps.org/doi/10.1103/PhysRevLett.123.163601}
}

@article{Cavalleri2010,
title = {Gas damping force noise on a macroscopic test body in an infinite gas reservoir},
journal = {Physics Letters A},
volume = {374},
number = {34},
pages = {3365-3369},
year = {2010},
issn = {0375-9601},
doi = {https://doi.org/10.1016/j.physleta.2010.06.041},
url = {https://www.sciencedirect.com/science/article/pii/S0375960110007279},
author = {A. Cavalleri and G. Ciani and R. Dolesi and M. Hueller and D. Nicolodi and D. Tombolato and S. Vitale and P.J. Wass and W.J. Weber},
}

@article{Giedke2001,
  title = {Entanglement Criteria for All Bipartite Gaussian States},
  author = {Giedke, G. and Kraus, B. and Lewenstein, M. and Cirac, J. I.},
  journal = {Phys. Rev. Lett.},
  volume = {87},
  issue = {16},
  pages = {167904},
  numpages = {4},
  year = {2001},
  month = {Oct},
  publisher = {American Physical Society},
  doi = {10.1103/PhysRevLett.87.167904},
  url = {https://link.aps.org/doi/10.1103/PhysRevLett.87.167904}
}

@book{GardinerZoller2010QuantumNoise,
  author    = {Gardiner, Crispin and Zoller, Peter},
  title     = {Quantum Noise: A Handbook of Markovian and Non-Markovian Quantum Stochastic Methods with Applications to Quantum Optics},
  edition   = {3},
  year      = {2010},
  month     = dec,
  publisher = {Springer},
  address   = {Berlin, Heidelberg},
  series    = {Springer Series in Synergetics},
  isbn      = {978-3-642-06094-6},
  pagetotal = {450},
  note      = {Softcover. Springer lists the hardcover (ISBN 978-3-540-22301-6) as published 27 Aug 2004; this softcover edition is published 01 Dec 2010.},
}

@article{kisil2021wienerhopf,
  author    = {Anastasia V. Kisil and I.D. Abrahams and G. Mishuris and S.V. Rogosin},
  title     = {The Wiener–Hopf technique, its generalizations and applications: constructive and approximate methods},
  journal   = {Proceedings of the Royal Society A},
  volume    = {477},
  number    = {20210533},
  year      = {2021},
  doi       = {10.1098/rspa.2021.0533},
  note      = {Modern survey of Wiener–Hopf factorization and applications}
}

@article{Vanner2011,
  author = {M. R. Vanner  and I. Pikovski  and G. D. Cole  and M. S. Kim  and \v{C}. Brukner  and K. Hammerer  and G. J. Milburn  and M. Aspelmeyer},
  title = {Pulsed quantum optomechanics},
  journal = {Proceedings of the National Academy of Sciences},
  volume = {108},
  number = {39},
  pages = {16182-16187},
  year = {2011},
  doi = {10.1073/pnas.1105098108},
  URL = {https://www.pnas.org/doi/abs/10.1073/pnas.1105098108}
}

@article{Hofer2011,
  title = {Quantum entanglement and teleportation in pulsed cavity optomechanics},
  author = {Hofer, Sebastian G. and Wieczorek, Witlef and Aspelmeyer, Markus and Hammerer, Klemens},
  journal = {Phys. Rev. A},
  volume = {84},
  issue = {5},
  pages = {052327},
  numpages = {10},
  year = {2011},
  month = {Nov},
  publisher = {American Physical Society},
  doi = {10.1103/PhysRevA.84.052327},
  url = {https://link.aps.org/doi/10.1103/PhysRevA.84.052327}
}

@article{Muller2008,
  title = {Entanglement of Macroscopic Test Masses and the Standard Quantum Limit in Laser Interferometry},
  author = {M\"uller-Ebhardt, Helge and Rehbein, Henning and Schnabel, Roman and Danzmann, Karsten and Chen, Yanbei},
  journal = {Phys. Rev. Lett.},
  volume = {100},
  issue = {1},
  pages = {013601},
  numpages = {4},
  year = {2008},
  month = {Jan},
  publisher = {American Physical Society},
  doi = {10.1103/PhysRevLett.100.013601},
  url = {https://link.aps.org/doi/10.1103/PhysRevLett.100.013601}
}

@article{Muller2009,
  title = {Quantum-state preparation and macroscopic entanglement in gravitational-wave detectors},
  author = {M\"uller-Ebhardt, Helge and Rehbein, Henning and Li, Chao and Mino, Yasushi and Somiya, Kentaro and Schnabel, Roman and Danzmann, Karsten and Chen, Yanbei},
  journal = {Phys. Rev. A},
  volume = {80},
  issue = {4},
  pages = {043802},
  numpages = {18},
  year = {2009},
  month = {Oct},
  publisher = {American Physical Society},
  doi = {10.1103/PhysRevA.80.043802},
  url = {https://link.aps.org/doi/10.1103/PhysRevA.80.043802}
}

@article{Yamamoto2006,
  title = {Robust observer for uncertain linear quantum systems},
  author = {Yamamoto, Naoki},
  journal = {Phys. Rev. A},
  volume = {74},
  issue = {3},
  pages = {032107},
  numpages = {10},
  year = {2006},
  month = {Sep},
  publisher = {American Physical Society},
  doi = {10.1103/PhysRevA.74.032107},
  url = {https://link.aps.org/doi/10.1103/PhysRevA.74.032107}
}

@article{Bouten2004,
  doi = {10.1088/0305-4470/37/9/010},
  url = {https://doi.org/10.1088/0305-4470/37/9/010},
  year = {2004},
  month = {feb},
  publisher = {},
  volume = {37},
  number = {9},
  pages = {3189},
  author = {Luc Bouten and Madalin Guta and Hans Maassen},
  title = {Stochastic Schrödinger equations},
  journal = {Journal of Physics A: Mathematical and General}
}

@article{Wieczorek2015,
  title = {Optimal State Estimation for Cavity Optomechanical Systems},
  author = {Wieczorek, Witlef and Hofer, Sebastian G. and Hoelscher-Obermaier, Jason and Riedinger, Ralf and Hammerer, Klemens and Aspelmeyer, Markus},
  journal = {Phys. Rev. Lett.},
  volume = {114},
  issue = {22},
  pages = {223601},
  numpages = {6},
  year = {2015},
  month = {Jun},
  publisher = {American Physical Society},
  doi = {10.1103/PhysRevLett.114.223601},
  url = {https://link.aps.org/doi/10.1103/PhysRevLett.114.223601}
}

@article{Korppi2018,
  title        = {Stabilized entanglement of massive mechanical oscillators},
  author       = {C. F. Ockeloen-Korppi and E. Damsk\"{a}gg and J.-M. Pirkkalainen and M. Asjad and A. A. Clerk and F. Massel and M. J. Woolley and M. A. Sillanp\"{a}\"{a}},
  journal      = {Nature},
  volume       = {556},
  pages        = {478--482},
  year         = {2018},
  doi          = {10.1038/s41586-018-0038-x},
  url          = {https://doi.org/10.1038/s41586-018-0038-x}
}

@article{Kotler2021,
  author = {Shlomi Kotler  and Gabriel A. Peterson  and Ezad Shojaee  and Florent Lecocq  and Katarina Cicak  and Alex Kwiatkowski  and Shawn Geller  and Scott Glancy  and Emanuel Knill  and Raymond W. Simmonds  and José Aumentado  and John D. Teufel },
  title = {Direct observation of deterministic macroscopic entanglement},
  journal = {Science},
  volume = {372},
  number = {6542},
  pages = {622-625},
  year = {2021},
  doi = {10.1126/science.abf2998},
  URL = {https://www.science.org/doi/abs/10.1126/science.abf2998}
}

@article{Lepinay2021,
  author = {Laure Mercier de Lépinay  and Caspar F. Ockeloen-Korppi  and Matthew J. Woolley  and Mika A. Sillanpää },
  title = {Quantum mechanics–free subsystem with mechanical oscillators},
  journal = {Science},
  volume = {372},
  number = {6542},
  pages = {625-629},
  year = {2021},
  doi = {10.1126/science.abf5389},
  URL = {https://www.science.org/doi/abs/10.1126/science.abf5389}
}
\end{document}